\documentclass[acmsmall, manuscript]{acmart}
\usepackage{booktabs}
\usepackage{multirow}
\usepackage{array}
\usepackage{longtable}
\usepackage{pdflscape} 
\usepackage{tikz}
\usetikzlibrary{positioning}
\usepackage[table]{xcolor}
\usepackage{hhline}

\graphicspath{{./}{figures/}}

\providecolor{BurntOrange}{RGB}{204,85,0}
\providecommand{\Description}[1]{}

\setcopyright{none}
\acmJournal{CSUR}
\acmYear{2026}
\copyrightyear{2026}
\acmDOI{}

\begin{document}

\title[Adaptive Microservice Management: Taxonomy \& Future Directions]{Adaptive Management of Microservices in Dynamic Computing Environments: A Taxonomy and Future Directions}

\author{Ming Chen}
\orcid{0000-0002-6120-6630}
\email{mingc4@student.unimelb.edu.au}

\author{Muhammed Tawfiqul Islam}
\orcid{0000-0003-4922-7807}
\email{tawfiqul.islam@unimelb.edu.au}

\author{Maria Rodriguez Read}
\orcid{0000-0002-2831-8526}
\email{maria.read@unimelb.edu.au}

\author{Rajkumar Buyya}
\orcid{0000-0001-9754-6496}
\email{rbuyya@unimelb.edu.au}

\affiliation{%
  \institution{the Quantum Cloud Computing and Distributed Systems (qCLOUDS) Laboratory, School of Computing and Information Systems, The University of Melbourne}
  \city{Parkville}
  \state{Victoria}
  \postcode{3010}
  \country{Australia}
}



\renewcommand{\shortauthors}{Chen, Islam, Read, and Buyya}

\begin{abstract}
Microservice-based cloud applications face changing workloads, evolving request paths, variable network conditions, interference, and failures. These dynamics couple autoscaling, placement, routing, isolation, and remediation. The survey examines dynamics-aware adaptive management for microservices. Its taxonomy covers control locus, modeled dynamics, adaptation strategy, and evaluation evidence; objectives and telemetry are cross-cutting. A synthesis of 84 system entries and 13 evaluation artifacts shows that production dynamics are often partially modeled. Reported gains also depend on evaluation fidelity. Key future directions include cross-layer coordination, telemetry-to-control abstractions, safe learning-based control, and reproducible dynamic evaluation.
\end{abstract}

\ccsdesc[300]{General and reference~Surveys and overviews}
\ccsdesc[500]{Computer systems organization~Cloud computing}
\ccsdesc[500]{Software and its engineering~Distributed systems organizing principles}
\ccsdesc[500]{Software and its engineering~Software performance}
\ccsdesc[300]{Software and its engineering~Software architectures}

\keywords{microservices, Kubernetes, cloud-native, autoscaling, scheduling, service mesh, observability, adaptive management, dynamic environments, reproducibility}
\maketitle

\section{Introduction}
\label{sec:intro}

Microservices decompose monolithic applications into a set of independently developed and deployed services, typically communicating via remote procedure call (RPC) and message-based interactions \cite{Cerny_SOA_MICROservices17}. A broad systems-oriented overview of microservices is given by Dragoni et al. \cite{Dragoni_Microservices2017}. In production, 
many microservice deployments are operated as \emph{cloud-native} applications, that is, containerized and managed by an orchestration platform (often on Kubernetes) \cite{2016borg_k8s,kubernetes}. This shift improves modularity and operational agility, but it also amplifies \emph{dynamic effects} that were less visible in monoliths: request traffic is bursty and non-stationary, request paths evolve as features are rolled out, and performance becomes sensitive to noisy-neighbor interference and network variability \cite{tail_at_scale2013,BubbleUp_colocation,Bubble_flux,DeepDive,Cloud_White}. For example, even when an application's microservice containers are optimally deployed across a cloud--edge cluster, end-to-end performance remains sensitive to time-varying cross-node delays \cite{TraDE_TPDS2026}. In this survey, \emph{adaptive management} refers to an online feedback loop that uses runtime telemetry to select actuation actions (e.g., resource allocation, placement, routing, or mitigation). We use the term \emph{dynamics} to denote time-varying behaviors that significantly affect microservice performance, cost, or reliability. We use \emph{dynamics-aware} to describe approaches that explicitly model or evaluate such behaviors, including workload bursts and shifts, evolving request paths and call graphs, network variability, interference, failures, and sustainability signals. These characteristics place microservice management in the class of partially observable, non-stationary control problems.

Modern cloud-native stacks provide powerful runtime control primitives (e.g., autoscaling, rescheduling, and traffic shifting), but using them effectively requires \emph{closing the control loop} under partial observability and evolving dynamics. Kubernetes exposes reactive controllers such as the Horizontal Pod Autoscaler (HPA) \cite{K8s_HPA} and Vertical Pod Autoscaler (VPA) \cite{k8s_vpa}, and is increasingly complemented by node- and resource-level controllers (e.g., cluster autoscaling, vertical autoscaling, and event-driven autoscaling) \cite{k8s_cluster_autoscaler,k8s_vpa,keda,knative,karpenter}. At the same time, production operators rely on observability pipelines to provide the telemetry needed for diagnosis and control, including metrics (Prometheus \cite{Prometheus}, Grafana \cite{grafana}), distributed tracing (Dapper \cite{dapper_tracing2010}, Jaeger \cite{jaeger}, Zipkin \cite{zipkin}), and open standards such as OpenTelemetry \cite{opentelemetry}. Service meshes and sidecar proxies provide additional control points for routing and latency-aware traffic management, including Istio \cite{istio}, Envoy \cite{Envoy}, and Linkerd \cite{linkerd}.

These capabilities motivate a growing body of research on \emph{adaptive management} for microservices. However, compared with classic cloud autoscaling or cluster scheduling, microservices introduce several distinctive challenges: (i) an application's decomposed services form a \emph{dependency graph} rather than a set of independent tasks \cite{Sage_ASPLOS21,Parslo_SoCC21,ChainFormer_ICSOC23,deployment_routing_TPDS23}; (ii) decisions at different layers (virtual machine (VM), container, call graph, critical path, and service) interact across timescales \cite{Borg_clusterManager,Fuxi_alibaba,k8s_network,TraDE_TPDS2026,FIRM_MS,GrahGeneration_ICS25}; and (iii) realistic performance evaluation requires capturing both system dynamics and orchestration dynamics under representative benchmarks and traces \cite{deathStarBench_ASPLOS19,muBench_TPDS23,trainTicket,github_AlibabaTraces,metaTraces_ATC23,k8sLoop_simu_PerformanceTools2024,chen2025idynamics}. Thus, this paper surveys \emph{dynamics-aware adaptive management} for microservices: approaches that explicitly account for time-varying behaviors and close the loop among observation, decision-making, and actuation. We adopt a systems perspective, focusing on modeling and control decisions that (i) improve service-level objective (SLO) satisfaction and efficiency under dynamic conditions and (ii) remain credible under realistic evaluation environments.

\subsection{Relation to Existing Surveys}
\label{sec:method}

\begin{table}[t]
\centering
\small
\renewcommand{\arraystretch}{1.2}
\begin{tabular}{p{0.15\linewidth} p{0.32\linewidth} p{0.46\linewidth}}
\hline
\textbf{Survey} & \textbf{Primary focus} & \textbf{Typical limitations}\\
\hline
\cite{senjab2023survey,k8s_scheduling_CSUR2022,k8s_CSUR} & Kubernetes scheduling and orchestration & Limited explicit treatment of call-graph dynamics and end-to-end SLOs; cross-layer coupling often remains implicit.\\
\cite{RootCauseAnalysis_CSUR_2025} & Root-cause localization in microservice systems & Limited discussion of infrastructure-level network variability; cross-layer coupling often remains implicit.\\
\cite{AlQassem_TNSM24} & Resource management for shared microservices & Emphasis on mechanisms; treatment of dynamics and evaluation environments is less systematic.\\
\cite{ml_resMam_Raj,ML_containerOrchestration} & Machine-learning-based resource management and orchestration & Often technology-centric; less emphasis on microservice dependency graphs and network dynamics.\\
\cite{perf_aware_cloudRes_taxonomy} & Performance-aware cloud resource management & Broad cloud perspective; microservice-specific request paths and service meshes are not central.\\
\cite{carbon_aware_yang2025survey} & Carbon- and energy-aware management & Focus on sustainability objectives; microservice-specific dynamics and orchestration constraints are less detailed.\\
\cite{eismann2020review_serverless} & Serverless platforms and elasticity & Different abstraction level (functions); the insights are useful but do not directly address microservice call graphs.\\
\cite{k8sLoop_simu_PerformanceTools2024} & Kubernetes-in-the-loop simulation and tools & Strong emphasis on evaluation realism; does not provide a full taxonomy of dynamics and adaptation mechanisms.\\
\hline
\textbf{This survey} & \multicolumn{2}{p{0.76\linewidth}}{Dynamics-aware microservice management: (1) an explicit taxonomy across multi-level dynamics, together with objectives and telemetry; and (2) a unified view spanning modeling, control, and evaluation.}\\
\hline
\end{tabular}
\caption{Representative surveys and their limitations from the perspective of dynamics-aware microservice management.}
\label{tab:survey-comparison}
\end{table}

\textit{How This Survey Differs from Existing Surveys.}
A substantial number of surveys cover parts of the cloud-native management stack, including Kubernetes scheduling, container orchestration, autoscaling, root-cause analysis, and cloud resource management. Examples include surveys on Kubernetes scheduling and orchestration \cite{senjab2023survey,k8s_scheduling_CSUR2022,k8s_CSUR}, containerized microservice resource-management frameworks \cite{AlQassem_TNSM24,sharedMS_AcmTranCS}, machine-learning-based resource management \cite{ml_resMam_Raj,ML_containerOrchestration}, performance-aware cloud resource management \cite{perf_aware_cloudRes_taxonomy}, and emerging topics such as carbon-aware management \cite{carbon_aware_yang2025survey}. In parallel, surveys of serverless computing provide complementary perspectives on elasticity and multi-tenant platforms \cite{eismann2020review_serverless}.

\textit{What Is Missing for Dynamics-Aware Microservice Management?}
While these surveys provide valuable foundations, three recurring gaps motivate a dedicated taxonomy for dynamics-aware microservice management.
\textbf{(1) Dynamics are simplified or treated implicitly.} Many surveys emphasize mechanisms (e.g., schedulers and autoscalers) without explicitly characterizing the \emph{types} of dynamics they address (e.g., workload, call-graph, network, and interference dynamics) or the interactions among them. For example, Kubernetes scheduling surveys \cite{senjab2023survey,k8s_scheduling_CSUR2022,k8s_CSUR} typically detail scheduling policies and extensibility points, but offer limited structure for classifying \emph{dynamic} network effects and evolving application dependencies.
\textbf{(2) Cross-layer interactions are underemphasized.} Microservice management spans multiple abstraction levels, including host/VM, container/pod, service, and call-graph relationships among upstream and downstream services. Surveys centered on a single layer (e.g., the Kubernetes scheduler) often overlook cross-layer feedback loops, such as ``autoscaler~$\leftrightarrow$~scheduler'' and ``traffic routing~$\leftrightarrow$~placement'' coupling \cite{TraDE_TPDS2026,Concury_loadBalancer,OptTraffic_ICPP23,collaborativeMesh_TSC2025}.
\textbf{(3) Evaluation realism and reproducibility remain fragmented.} Many studies still rely on synthetic workloads or simplified simulators, whereas production environments increasingly exhibit realistic request-path evolution and orchestration behavior \cite{deathStarBench_ASPLOS19,muBench_TPDS23,trainTicket,github_AlibabaTraces,metaTraces_ATC23}. Recent work, therefore, highlights the need for \emph{Kubernetes-in-the-loop} simulation and emulation to narrow the realism gap \cite{k8sLoop_simu_PerformanceTools2024,Kollaps_EuroSys20,THUNDERSTORM_SRDS19,chen2025idynamics}.

\textit{Survey Comparison.}
Table~\ref{tab:survey-comparison} summarizes representative surveys and the aspects they emphasize. In contrast, this paper explicitly structures the literature around (i) \emph{where control is applied}, (ii) \emph{which dynamics are modeled}, (iii) \emph{how adaptation is performed}, and (iv) \emph{how systems are evaluated}. Our contribution is not the claim that prior subareas lack taxonomies; rather, it is a unified multi-dimensional synthesis that connects control placement, explicit dynamics classes, adaptation strategy, and evaluation fidelity across previously separate literatures.

\subsection{Scope, Corpus, and Review Methodology}
\label{sec:scope-corpus}

\textit{Scope.} This survey focuses on runtime adaptive management for microservice-based cloud applications: work in which online telemetry informs actuation decisions that affect SLO satisfaction, cost, efficiency, reliability, or sustainability under time-varying conditions. We include papers whose central contribution is a controller, model, or management mechanism operating at the host/VM, cluster, service, application-graph, service-mesh, network, or edge level. We also include adjacent systems work when it directly informs microservice management along one of our taxonomy axes---for example, datacenter transport and isolation papers that shape network or interference control, or cloud--edge placement papers that expose dynamics not visible within a single cluster. We exclude work focused purely on design-time decomposition, static architecture refactoring, security, or policy enforcement without runtime adaptation, or serverless-only control that does not expose microservice dependency-graph behavior. Evaluation artifacts (benchmarks, traces, emulators, simulators, and Kubernetes-in-the-loop tools) are analyzed separately in Table~\ref{tab:benchmarks-traces} rather than being mixed into the system-comparison corpus.

\textit{Corpus construction.} Because the literature is fragmented across systems, networking, cloud, performance engineering, and AIOps venues, we adopt a structured narrative-survey process rather than claiming an exhaustive bibliometric analysis. The corpus was assembled in three passes. First, we seeded the review from representative prior surveys (Table~\ref{tab:survey-comparison}), foundational platform primitives, and canonical systems papers that recur in discussions of microservice management. Second, we expanded this seed set by backward and forward snowballing over highly relevant papers and artifacts, following common evidence-synthesis and mapping practices in software engineering \cite{Wohlin2014,Petersen2015}. Third, we performed venue- and keyword-guided screening to capture work on autoscaling, scheduling, placement, traffic management, isolation, diagnosis, remediation, and evaluation tooling in cloud-native microservice settings through early~2026. This process prioritizes archival systems, networking venues, and journals, while allowing a small number of practitioner or platform-documentation sources when the goal is to describe deployed control primitives or public artifacts rather than to claim empirical superiority.

\textit{Coded representative corpus.} The main coded corpus used in Section~\ref{sec:taxonomy} contains 84 representative system entries: 75 research papers and 9 documented production primitives or platform systems. These are complemented by 13 evaluation artifacts in Table~\ref{tab:benchmarks-traces}, which are analyzed separately because they serve as evidence infrastructure rather than as controllers. Each system entry is coded along D1--D4 and then used in two ways: first, to support the comparative synthesis in Table~\ref{tab:comparative-synthesis}; and second, to support the descriptive corpus-level observations in Section~\ref{subsec:eval-repro}. The counts reported there should therefore be read as descriptive statistics for a curated representative corpus, not as claims about the exact prevalence of every technique in the full literature. This distinction is important for claim calibration: our goal is to expose structural patterns, recurrent evaluation habits, and underexplored combinations, not to present a formal meta-analysis of effect sizes across incomparable workloads, SLO definitions, and testbeds.

\subsection{Contributions of This Survey}
\label{sec:contributions}

This survey makes four main contributions.

\begin{itemize}
  \item \textbf{Unified taxonomy for dynamics-aware microservice management.} We propose a taxonomy that separates \emph{dynamics} from \emph{mechanisms}, elevates the \emph{evaluation environment} to a primary dimension, and organizes the literature by \emph{where} control is applied (D1), \emph{which} dynamics are modeled (D2), \emph{how} adaptation is performed (D3), and \emph{how} systems are evaluated (D4), while treating objectives and telemetry as cross-cutting dimensions.
  \item \textbf{Critical cross-literature comparative synthesis.} Using this taxonomy, we connect previously separate literatures and compare representative approaches across graph-aware autoscaling, scheduling, and placement, traffic management, isolation, diagnosis, and remediation in terms of modeling assumptions, observability requirements, actuators, controller interactions, and evaluation fidelity.
  \item \textbf{Evaluation and reproducibility guidance for dynamic environments.} We consolidate commonly used benchmarks, traces, load generators, emulators, simulators, in-the-loop tools, and reproducibility practices, and distill practical guidance on scenario matrices, baselines, metrics, ablations, and artifact reporting for credible comparison.
  \item \textbf{Research gaps and future directions.} We identify open problems in cross-layer controller coordination and stability, telemetry-to-control under partial observability, coupled multi-dynamics and multi-objective control, safe and trustworthy learning-based controllers, standardized dynamic evaluation pipelines, and the emerging role of agentic and multi-agent LLMs as supervisory layers for microservice management.
\end{itemize}

\subsection{Organization}
\label{sec:organization}
The remainder of the survey is organized as follows. Section~\ref{sec:background} provides background on cloud-native microservices, orchestration platforms, service meshes, observability pipelines, the primary sources of runtime dynamics, and the evaluation artifacts used in the literature. Section~\ref{sec:taxonomy} motivates the taxonomy and proposes the classification used throughout the survey. Within Section~\ref{sec:taxonomy}, Section~\ref{subsec:taxonomy-overview} summarizes the overall design space; Sections~\ref{subsec:D1}–\ref{subsec:D4} discuss the four primary dimensions (D1–D4); Section~\ref{subsec:crosscutting} examines the cross-cutting dimensions of objectives/SLOs and telemetry; and Section~\ref{subsec:eval-repro} consolidates comparative evaluation methodologies and reproducibility guidance. Section~\ref{sec:challenges} discusses open challenges and future directions, including cross-layer coordination, telemetry-to-control under partial observability, coupled multi-dynamics and multi-objective control, trustworthy learning-based controllers, standardized dynamic scenarios and reproducible evaluation pipelines, and agentic and multi-agent LLM-based management. Finally, Section~\ref{sec:conclusion} summarizes the paper and offers concluding remarks.

\section{Background: Cloud Microservices, Runtime Dynamics, and Evaluation Artifacts}
\label{sec:background}

\subsection{Cloud Microservices and Orchestration Platforms}
Microservices are typically deployed as containers and managed by an orchestration platform that provides:
(i) resource abstraction (pods, services, namespaces),
(ii) placement and scheduling,
(iii) elastic scaling, and
(iv) fault recovery and rollout primitives \cite{kubernetes,2016borg_k8s}.
Kubernetes, inspired by earlier cluster managers such as Borg and by extensible shared-state control-plane designs such as Mesos and Omega \cite{Borg_clusterManager,2016borg_k8s,Mesos_NSDI11,Omega_EuroSys13}, is among the most widely adopted orchestrators for containerized microservices, including in hybrid cloud--edge settings \cite{goudarzi2022fogScheduling,k8s_fog_latency,adaptive_ms_cloudEdge_TPDS21}.

\textbf{Autoscaling and Control Loops.}
At the pod level, Kubernetes provides reactive horizontal scaling (HPA) \cite{K8s_HPA}.
In practice, production deployments often use additional controllers: vertical scaling (VPA) to adjust resource requests and limits \cite{k8s_vpa}, cluster autoscaling to add or remove nodes \cite{k8s_cluster_autoscaler,karpenter}, and event-driven scaling to handle bursty or queue-based workloads \cite{keda,knative}.
This naturally leads to \emph{multi-loop} control: scaling decisions interact with service placement/scheduling and can create oscillation or delayed convergence if not coordinated \cite{k8s_interference_TCC,Dirigent_LC_QoS}.

\textbf{Service Meshes and Traffic Management.}
Many service meshes (e.g., Istio/Envoy/Linkerd) implement routing, telemetry, and policy enforcement via sidecar proxies \cite{istio,Envoy,linkerd}, but recent designs also explore sidecar-free and multi-tenant dataplanes such as Canal Mesh \cite{CanalMesh_SIGCOMM24}.
They enable fine-grained traffic shifting, retries, and circuit breaking for SLO management or joint placement-routing control, while recent empirical and testing work shows that mesh policy behavior itself must be evaluated explicitly \cite{Concury_loadBalancer,collaborativeMesh_TSC2025,collaborativeMesh_Infocom25,OptTraffic_ICPP23,ServiceMeshPolicies_ICPE22,MeshTest_NSDI25}.

\textbf{Observability Pipelines.}
Microservice operation depends on telemetry to diagnose and adapt to changing conditions.
Metrics systems (e.g., Prometheus) \cite{Prometheus} provide low-cost time-series signals; distributed tracing systems (e.g., Dapper/Jaeger/Zipkin) \cite{dapper_tracing2010,jaeger,zipkin} reveal request paths and latency breakdowns; dashboards (Grafana) \cite{grafana} support exploratory analysis; and OpenTelemetry provides vendor-neutral instrumentation and export for metrics/logs/traces \cite{opentelemetry}.
Beyond tooling, classic and recent tracing systems such as X-Trace, Pivot Tracing, tprof, STEAM, and DeepFlow show that control-relevant telemetry must preserve causality and observability under overhead and sampling constraints \cite{XTrace_NSDI07,PivotTracing_SOSP15,TProf_SoCC21,STEAM_FSE23,DeepFlow_SIGCOMM23}.
Observability signals are therefore both an \emph{input} to controllers and a \emph{measurement} channel for evaluation (e.g., SLO violations).

\begin{figure*}[t]
  \centering
  \includegraphics[width=\textwidth]{figures/four_dominant_dynamics.png}
  \caption{Overview illustration of four common origins of microservice dynamics: demand-side variation, application and configuration evolution, shared-resource contention, and network and infrastructure variability. The center depicts a representative service graph, and surrounding arrows indicate that these origins manifest as workload changes, call-graph evolution, contention, and network traffic variation.}
  \Description{Overview diagram showing four origins of microservice dynamics: demand-side variation, application and configuration evolution, shared-resource contention, and network and infrastructure variability. The center depicts a representative service graph, and surrounding arrows indicate that these origins manifest as workload changes, call-graph evolution, contention, and network traffic variation.}
  \label{fig:four-dynamics-origins}
\end{figure*}

\subsection{Primary Sources of Runtime Dynamics}


Operational dynamics in microservice systems arise primarily from four origins:
(i) \textbf{demand-side variation}, including flash crowds, diurnal cycles, bot-driven
request spikes, and tenant onboarding;
(ii) \textbf{application and configuration evolution}, including canary releases,
feature flags, API version migrations, schema changes, and routing-policy updates;
(iii) \textbf{shared-resource contention and multi-tenancy}, including noisy-neighbor
interference on CPU, memory, storage, network, and accelerators; and
(iv) \textbf{network and infrastructure variability}, including cross-zone latency changes,
transient packet loss or queueing, node failures, rescheduling, rolling upgrades, and
autoscaler-driven rebalancing.

These origins manifest as several recurring dynamic classes in the literature:
workload/resource-usage dynamics, call-graph and communication-mode dynamics,
network/traffic dynamics, contention/interference dynamics, failure/quality of service (QoS)-degradation
dynamics, and sustainability-related signals or constraints. We separate
\emph{origins} from \emph{manifestations}: the former describe where change comes from,
while the latter describes how it appears in the running system.

\begin{itemize}
    \item \textbf{Workload and resource-usage dynamics.}
Cloud workloads are well known to be bursty and heavy-tailed, with non-stationary patterns across seconds to days \cite{Bitbrains,Alibaba_colocatedData,Borg_trace,MLaas_Alibaba_GPU_traces}.
Workload prediction is therefore widely used in proactive scaling and capacity planning.
Classic statistical methods (e.g., ARIMA) \cite{arima_ref1,Arima_ref2,Arima_ref3,arimaPred_raj} are still competitive for certain regimes, while deep learning approaches (LSTM/TCN/Transformer variants) improve robustness under non-linearity and long-range dependencies \cite{LSTM_baseline,LSTM_worklod,lstm_workloadPred,DeepTCN,esDNN_minxian,GNN_timeSeries,chen2023enbeats}.
In microservices, workload dynamics propagate through the dependency graph: an upstream burst can amplify downstream contention, causing tail-latency spikes \cite{tail_at_scale2013,Sage_ASPLOS21,imporvedQueueNetworks}.

    \item \textbf{Application, call-graph, and communication-mode dynamics.}
Unlike monoliths, microservices often change their request paths due to feature rollout, canary releases, A/B testing, and adaptive routing \cite{Sage_ASPLOS21,Parslo_SoCC21,ChainFormer_ICSOC23,deployment_routing_TPDS23,GrahGeneration_ICS25}.
These changes lead to \emph{call-graph dynamics} (which services are involved and how frequently) and \emph{communication-mode dynamics} (e.g., synchronous RPC vs.\ async messaging, fan-out patterns) \cite{GrahGeneration_ICS25}.
Call-graph and dependency inference also underpins debugging and incident management \cite{Seer,ART_ASE24,PerfScope_bug_inference}.

    \item \textbf{Network and traffic dynamics.}
Network variability includes bandwidth contention, queuing in shared switches, cross-rack latency, and dynamic routing.
Such effects become more visible as microservices communicate frequently and may be distributed across zones or edge sites \cite{D2TCP_netBand_isolate,Fastpass_netBand_isolate,netBand_isolate,k8s_network,NetMARKS_InfoCOM21,ExtendingNetK8s_ICSOC22,ExtendingNetK8s_CCGRID22,OptTraffic_ICPP23}.
Recent work explores joint scheduling with in-depth traffic analysis to manage network dynamics and improve SLOs \cite{TraDE_TPDS2026,Concury_loadBalancer,collaborativeMesh_TSC2025}. Canonical datacenter traffic-management systems such as Ananta and Maglev show how software load balancing itself becomes a cloud-scale control primitive, while pFabric illustrates how flow-completion-sensitive transport can materially affect short RPC-like microservice traffic \cite{Ananta_SIGCOMM13,Maglev_NSDI16,pFabric_SIGCOMM13}.

    \item \textbf{Resource contention, co-location, and interference.}
Microservices are commonly \emph{shared} across tenants or co-located with batch jobs, leading to interference on CPU, memory, storage, and accelerators \cite{BubbleUp_colocation,Bubble_flux,DeepDive,CPI2_perfIsolation_cpu,Cloud_White,PIMCloud_LCjobs,PARTIES,CLITE_colcation_LCjobs}.
Both system-level mechanisms (isolation, bandwidth control) and scheduler-level policies are used to mitigate contention \cite{EyeQ_isolate_netBand,Bubble_flux,k8s_interference_TCC,Dirigent_LC_QoS}.
At the microservice layer, specialized resource managers aim to improve utilization while meeting service-level agreement (SLA)/SLO targets \cite{FIRM_MS,Erms_ASPLOS23,Sinan_MS,grandslam_MS,Derms_ISCA2024,sharedMS_AcmTranCS}.

\item \textbf{Additional dynamics: failures and sustainability.}
Dynamic environments also include failure events (node failures, stragglers, overload, QoS degradation) \cite{Fuxi_alibaba,Cloud_White,Seer}, and sustainability-related variability such as time-varying carbon intensity or energy budgets \cite{carbon_aware_yang2025survey,saxena2022proactive,Hipster_LC_task_manager}. Recent systems work shows both the promise of carbon-aware scheduling under SLO constraints and the practical limits of temporal or spatial workload shifting for latency-sensitive services \cite{CASPER_IGSC23,CarbonShiftLimit_EuroSys24}.
These dynamics motivate controllers that incorporate robustness, risk, and long-term objectives rather than short-term latency alone.

\end{itemize}

{\scriptsize
\begin{table}[t]
\centering
\small
\renewcommand{\arraystretch}{1.15}
\begin{tabular}{p{0.23\linewidth} p{0.2\linewidth} p{0.5\linewidth}}
\hline
\textbf{Evaluation artifacts} & \textbf{Type} & \textbf{Dynamics covered}\\
\hline
DeathStarBench \cite{deathStarBench_ASPLOS19} & benchmark suite & realistic microservice request paths; tail-latency sensitivity; resource contention.\\
$\mu$Bench \cite{muBench_TPDS23} & benchmark suite & microservice scaling and placement scenarios; configurable bottlenecks.\\
TrainTicket \cite{trainTicket} & benchmark app & realistic service graph and RPC patterns; supports load testing for scaling studies.\\
Alibaba traces \cite{github_AlibabaTraces,Alibaba_colocatedData} & production traces & non-stationary workload; co-location and interference; cluster-scale patterns.\\
Meta traces \cite{metaTraces_ATC23} & production traces & microservice-level request patterns and dependencies (trace-driven evaluation).\\
Bitbrains \cite{Bitbrains} & cloud traces & VM-level workload and resource-usage dynamics; used for forecasting baselines.\\
wrk2 \cite{wrk2} & load generator & controlled workload bursts; used in real experiments.\\
Locust \cite{locust} & load generator & scriptable user-level workload generation; used for microservice stress and scalability experiments.\\
JMeter \cite{jmeter} & load generator & protocol-level workload generation and performance testing; widely used in service performance evaluation.\\

Network emulation \cite{Kollaps_EuroSys20,THUNDERSTORM_SRDS19} & emulation & time-varying network delay/bandwidth; fault injection.\\
K8s-in-the-loop \cite{k8sLoop_simu_PerformanceTools2024} & simulation & simulated events; coupling among schedulers.\\
iDynamics \cite{chen2025idynamics} & evaluation framework & authentic orchestration control; emulated network; controllable dynamics; coupling scheduling policies.\\
MeshTest \cite{MeshTest_NSDI25} & evaluation framework & end-to-end validation of mesh traffic policies; routing-policy correctness and failure-triggering interactions.\\

\hline
\end{tabular}
\caption{Representative artifacts (benchmarks/traces/tools) for evaluating microservices under dynamics.}
\label{tab:benchmarks-traces}
\end{table}
}

\subsection{Evaluation Artifacts: Benchmarks, Traces, and Tooling for Evaluating Dynamics}
A recurring challenge in dynamics-aware microservice management is evaluation: synthetic workloads are easy to reproduce but often miss microservice coupling, network effects, or orchestration behavior.
In this survey, we use \emph{evaluation artifacts} to denote tangible and reusable evidence used to assess a system, including benchmark applications, workload generators, production traces, and simulators/emulators.
These evaluation artifacts are crucial and serve three complementary purposes: \textbf{validation} (confirming that a method works as claimed), \textbf{improvement} (enabling iterative refinement during system design), and \textbf{assessment} (judging final effectiveness under representative conditions).
They are also central to reproducibility, because they expose the assumptions, data, and execution environment required to replay results and compare approaches fairly.
Table~\ref{tab:benchmarks-traces} summarizes commonly used evaluation artifacts for dynamics-aware microservice studies and the types of dynamics they help reproduce.

\section{Taxonomy of Dynamics-Aware Cloud and Microservice Management}
\label{sec:taxonomy}
\subsection{Design Rationale}
The taxonomy is intended to serve as a comparative framework rather than a descriptive list of techniques. Its role is to expose the recurring dimensions along which prior work actually differs and along which transferability must be judged: where the control loop closes, which runtime dynamics are modeled, how decisions are produced, how those decisions are enacted, and what evaluation evidence supports the reported gains.

The taxonomy follows three principles. First, \emph{modeled dynamics} are kept separate from the \emph{adaptation strategy}, because the same dynamics class can be addressed by different controllers, and the same controller family can be reused across different dynamics. For example, the same adaptation strategy (e.g., scaling) can be used for resolving different dynamics (service-demand vs.\ resource contention), and the same dynamics (e.g., interference) can be handled by different mechanisms (isolation vs.\ scheduling). Second, the \emph{adaptation strategy} is decomposed into \emph{decision logic} (D3a) and \emph{actuation mechanism} (D3b). This split avoids conflating controller form with operational control knobs: predictive modeling, optimization, learning, and diagnosis describe how decisions are derived, whereas scaling, placement, routing, isolation, and remediation describe how those decisions are executed. Third, evaluation is represented as a structured dimension of \emph{evaluation evidence} (D4a--D4c), because the performance strength of a systems claim depends not only on the reported outcome, but also on the \emph{execution substrate}, the \emph{workload source}, and the presence or absence of \emph{fidelity enhancers} such as network emulation, fault injection, or Kubernetes-in-the-loop execution \cite{Kollaps_EuroSys20, THUNDERSTORM_SRDS19, chen2025idynamics, k8sLoop_simu_PerformanceTools2024}.

The proposed taxonomy is permissive about overlap. A representative system may occupy multiple branches in D1--D3 when control spans several loci, models coupled dynamics, or combines multiple actuators. By contrast, D4 is recorded as \emph{evaluation evidence}, since each study necessarily makes a concrete evidentiary commitment even when that commitment is not presented as a formal design choice. Specific taxonomies and relations between D1, D2, D3, and D4 dynamics dimensions can be found in Figure~\ref{fig:taxonomy-diagram}.

\subsection{Taxonomy Overview}
\label{subsec:taxonomy-overview}
\begin{figure}[t]
        \centering
        \includegraphics[width=\linewidth]{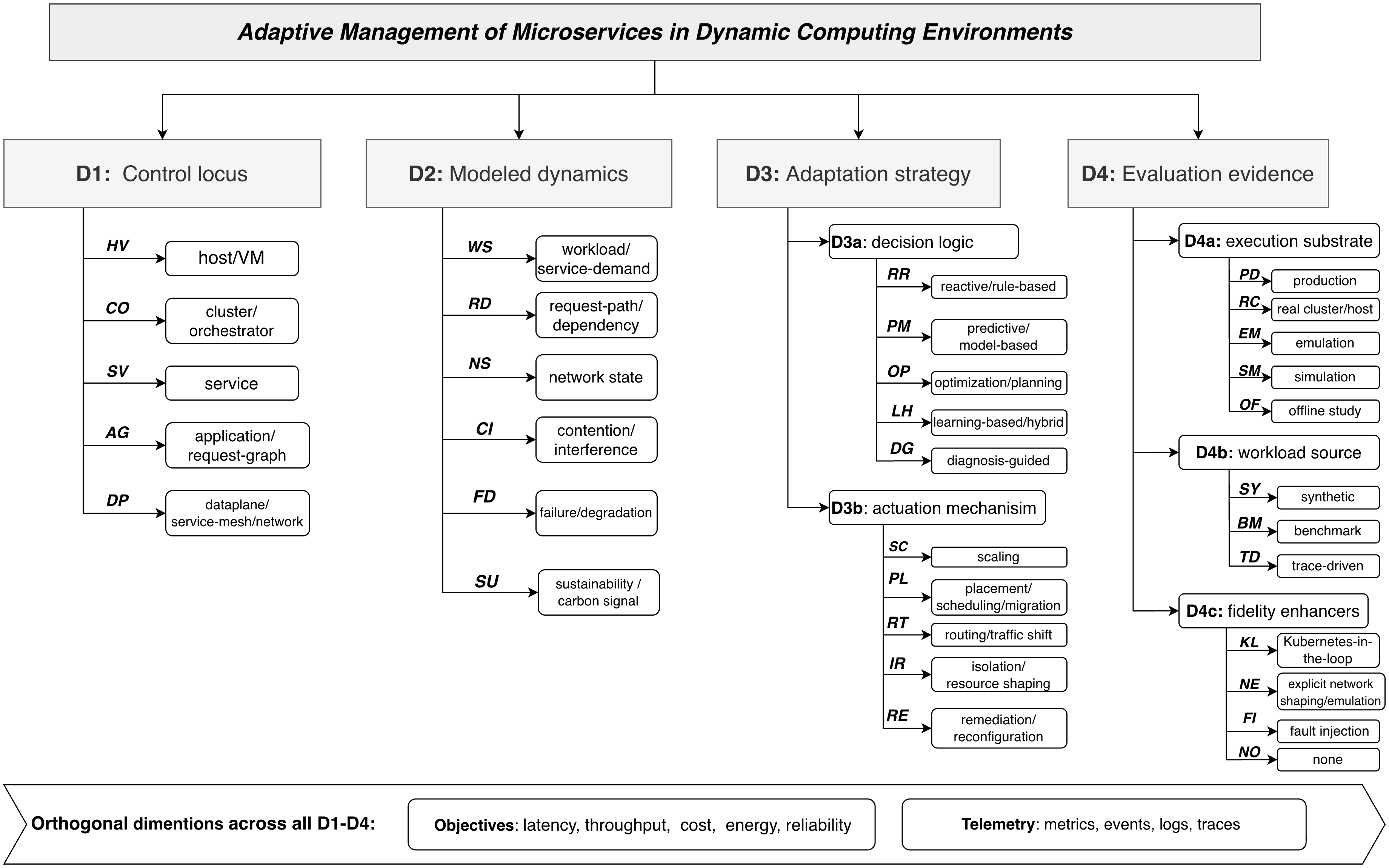}
        \caption{The proposed taxonomy for dynamics-aware microservice management. D1 records control locus; D2 records modeled dynamics; D3 separates decision logic (D3a) from actuation mechanism (D3b); and D4 records evaluation evidence through execution substrate (D4a), workload source (D4b), and fidelity enhancers (D4c). Objectives/SLOs and telemetry remain cross-cutting dimensions.}
  \Description{The proposed taxonomy diagram for dynamics-aware microservice management. The primary branches encode D1 control locus, D2 modeled dynamics, D3 adaptation strategy split into decision logic and actuation mechanism, and D4 evaluation evidence split into execution substrate, workload source, and fidelity enhancers. Objectives/SLOs and telemetry are shown as cross-cutting dimensions.}
        \label{fig:taxonomy-diagram}
\end{figure}

The four primary dimensions (D1--D4) in Figure~\ref{fig:taxonomy-diagram} provide the main classification used throughout this survey. D1 records the \emph{control locus}---where control decisions are applied in the system stack; D2 records the \emph{modeled dynamics}---which types of dynamics are explicitly modeled; D3 records \emph{adaptation strategy}---how adaptation is realized through models, controllers, and actuators---via two linked sub-dimensions: decision logic (D3a) and actuation mechanism (D3b); and D4 records \emph{evaluation evidence}---how proposed systems are evaluated---via execution substrate (D4a), workload source (D4b), and fidelity enhancers (D4c). This organization is deliberate: the same application-graph controller may rely on predictive modeling, optimization, or learning; the same routing or scaling action may be applied at different control loci; and apparently strong gains may weaken once the supporting evaluation evidence becomes more realistic.


Figure~\ref{fig:taxonomy-diagram} should therefore be read as a classification
schema rather than as a strict pipeline. A single system may span several
branches in D1--D3, but nearly every study makes an implicit or explicit
commitment along all four dimensions. For compact comparison,
Table~\ref{tab:taxonomy-d1d4} uses abbreviated codes for the D1--D4
subcategories, and Table~\ref{tab:eval-patterns} aggregates selected
patterns from those codes. Multi-label entries indicate that more than one
locus, dynamics class, decision logic, actuator, or evidence mode is explicit
in the cited work. The full code legend is provided with
Table~\ref{tab:taxonomy-d1d4}; Table~\ref{tab:eval-patterns} includes a
short note for the subset of codes used in the aggregate counts.

In addition, two \emph{orthogonal dimensions} repeatedly shape design choices across D1--D4: \emph{objectives/SLOs} (e.g., tail-latency, cost, energy, and reliability) and \emph{telemetry} (metrics, logs, traces, and related runtime signals). We treat them as cross-cutting rather than primary axes because they constrain every stage of the control loop: objectives determine what the system seeks to optimize, while telemetry determines what can be observed, inferred, and acted upon. The following subsections therefore examine D1--D4 individually, and then return to these two cross-cutting axes together with comparative evaluation and reproducibility guidance.

\subsection{Dimension 1: Control Locus}
\label{subsec:D1}

D1 identifies \emph{where} a feedback loop closes. Adaptive microservice-management decisions can be applied at different, non-mutually exclusive abstraction levels. We distinguish the following common control loci.

\textbf{HV: Host/VM.} Controls at this locus manage physical servers or virtual machines through provisioning, consolidation, and performance isolation. Representative systems include interference and QoS managers for co-located latency-critical workloads, such as Bubble-Up/Bubble-Flux, PARTIES, CPI$^2$, Heracles, PIMCloud, and Dirigent \cite{BubbleUp_colocation,Bubble_flux,PARTIES,CPI2_perfIsolation_cpu,Heracles,PIMCloud_LCjobs,Dirigent_LC_QoS}, as well as VM-level anomaly prediction and prevention mechanisms such as PREPARE \cite{PREPARE_perfPredict}. This locus is attractive because it requires no application changes, but its effect on end-to-end SLOs is usually indirect and often mediated through higher-layer service behavior.

\textbf{CO: Cluster-orchestrator.} This locus covers cluster scheduling, node provisioning, autoscaling coordination, and controller interaction in systems such as Kubernetes. Research in this area includes scheduler extensions and network-aware placement policies \cite{k8s_network,NetMARKS_InfoCOM21,ExtendingNetK8s_ICSOC22,ExtendingNetK8s_CCGRID22,k8s_optim_ms}, together with QoS-aware and heterogeneity-aware cluster management exemplified by large-scale schedulers such as Borg and Fuxi and research systems such as Quasar and Paragon \cite{Borg_clusterManager,Fuxi_alibaba,Quasar_QoS_aware_clusterManagement,Paragon_QoS_hetero}. CO exposes strong actuators---especially scheduling, migration, and node scale-out/in---but it also inherits cross-controller coupling and can only exploit dynamics that the orchestrator actually observes.

\textbf{SV: Service.}
Service-level control applies decisions to an individual microservice or replica, often targeting SLO/SLA satisfaction with improved utilization.
Representative systems include FIRM, Sinan, GrandSLAm, and ERMS \cite{FIRM_MS,Sinan_MS,grandslam_MS,Erms_ASPLOS23}, and recent designs for shared microservices and scalable SLA management \cite{sharedMS_AcmTranCS,AlQassem_TNSM24,Derms_ISCA2024}. Beyond hard resources such as CPU and memory, some approaches adapt \emph{soft resources} such as concurrency limits on threads or database connections (e.g., $\mu$ConAdapter) \cite{ms_rl_SOCC23}, and multifaceted reinforcement-learning (RL) scaling frameworks (e.g., CoScal) combine multiple control knobs \cite{ms_rl_TNSM}. Safe learning-based designs aim to reduce exploration risk under rare events \cite{SafeDRL_TC24}. Thus, per-service control offers fine-grained leverage, but it must account for dependency structure to avoid local decisions that degrade end-to-end performance.

\textbf{AG: Application/request graph.}
Here, the unit of management is a \emph{request path} (e.g., critical path) or service dependency graph, and the goal is end-to-end performance across services.
Trace studies and datasets characterize dependency structures and workflow variability in production microservice deployments
\cite{SoCC2021_MS_luo,metaTraces_ATC23,github_AlibabaTraces}.
Systems such as Parslo and ChainsFormer explicitly model microservice graphs to allocate SLO budgets or scale critical chains \cite{Parslo_SoCC21,ChainFormer_ICSOC23},
while Sage and Seer leverage telemetry to diagnose dependency-induced QoS issues \cite{Sage_ASPLOS21,Seer}.
Graph-aware service placement and routing further combine dependency and network effects \cite{deployment_routing_TPDS23,TraDE_TPDS2026,Net_ms_placement_TNSM},
and graph generation supports scalable what-if evaluation \cite{GrahGeneration_ICS25}. Therefore, AG can be more directly aligned with end-to-end SLOs when dependency structure and request-path telemetry are available, but it depends heavily on high-quality tracing and can be sensitive to partial observability.

\textbf{DP: Dataplane/service-mesh/network.} This locus captures request steering, transport behavior, sidecar or proxy policies, and other dataplane controls that directly shape communication behavior at runtime. Examples include software load balancers and service-mesh routing systems \cite{Concury_loadBalancer,collaborativeMesh_Infocom25,collaborativeMesh_TSC2025,CanalMesh_SIGCOMM24,istio,Envoy,linkerd}, as well as network-isolation and transport mechanisms that motivate microservice-aware dataplane reasoning more broadly \cite{D2TCP_netBand_isolate,Fastpass_netBand_isolate,EyeQ_isolate_netBand,netBand_isolate}. DP is often the fastest control locus, but it is also the easiest place to create harmful interactions with autoscaling, placement, and retry policies if cross-layer coordination is missing.

\textbf{Deployment-span modifier \texttt{[ED]}.} When a work explicitly spans cloud--edge, multi-cluster, or geo-distributed settings, Table~\ref{tab:taxonomy-d1d4} appends an \texttt{[ED]} modifier to the dominant control locus instead of introducing another locus in D1. This keeps the ontology parallel: edge deployment changes the \emph{span} of a controller, but the controller still closes primarily at a service, graph, cluster-orchestrator, or dataplane locus. Representative work includes cloud--edge placement and migration \cite{adaptive_ms_cloudEdge_TPDS21,Adaptive_ms_IPDPS21,delay_edge_SPE22,delay_ms_TMC22}, reliability-aware fog placement \cite{Net_ms_placement_TNSM,ms_placement_iot_TMC,Pallewatta_CSUR23}, and edge-oriented Kubernetes orchestration \cite{k8s_fog_latency}.

\textbf{Takeaway of (D1).}
D1 captures \emph{where} the control loop closes. Moving from HV/CO toward SV/AG often improves SLO alignment and application-level semantic awareness, whereas DP typically offers the fastest reaction path. In practice, effective deployment-ready designs require richer telemetry and stronger cross-layer coordination.

\subsection{Dimension 2: Modeled Dynamics}
\label{subsec:D2}
D2 records the time-varying phenomena that are modeled explicitly in the controller state, performance model, or reasoning process. Dynamics-aware management systems differ primarily in \emph{which} dynamics they represent and respond to.

\textbf{WS: Workload/service-demand dynamics.} WS captures variation in arrival rate, request mix, service demand, queue growth, and multi-resource consumption. Many proactive approaches forecast workload or service-demand using statistical and learning-based models
\cite{arimaPred_raj,CloudInsight_wokrloadPre,lstm_workloadPred,DeepTCN,esDNN_minxian,chen2023enbeats},
and translate predictions into provisioning or scaling decisions.
Complementarily, analytical and hybrid performance models (e.g., queueing-based predictors) connect demand to latency and can support SLO planning
\cite{imporvedQueueNetworks,PREPARE_perfPredict,deployment_routing_TPDS23}.
These dynamics are often non-stationary (bursts, diurnal cycles, workload shifts), which motivates robust online adaptation rather than static tuning.

\textbf{RD: Request-path/dependency dynamics.} RD captures changes in service dependencies, feature rollouts, critical paths, fan-out structure, traffic shifts, and communication patterns across a microservice graph. Trace studies provide empirical evidence of service dependency variability and performance heterogeneity in large deployments \cite{SoCC2021_MS_luo,metaTraces_ATC23,github_AlibabaTraces},
motivating methods that explicitly model microservice call-graphs.
Parslo and ChainsFormer incorporate the dependency structure into SLO allocation and scaling decisions \cite{Parslo_SoCC21,ChainFormer_ICSOC23},
while diagnosis systems such as Sage and Seer leverage telemetry to reason about dependency-induced QoS violations \cite{Sage_ASPLOS21,Seer}.
At the actuation layer, deployment and routing designs adapt to changing dependencies and critical paths \cite{deployment_routing_TPDS23,TraDE_TPDS2026},
and graph generators help evaluate controllers under diverse evolving workflows \cite{GrahGeneration_ICS25}.

\textbf{NS: Network-state dynamics.} NS captures cross-node latency, bandwidth variation, topology locality, traffic imbalance, and communication-path state that affect microservice performance over time. Classic datacenter transport and bandwidth control mechanisms address flow-level dynamics \cite{D2TCP_netBand_isolate,Fastpass_netBand_isolate,netBand_isolate},
while recent cloud-native work integrates network signals into orchestration and scheduling
\cite{NetMARKS_InfoCOM21,k8s_network,ExtendingNetK8s_ICSOC22,ExtendingNetK8s_CCGRID22,k8s_fog_latency}.
On the traffic-management side, load balancers and service meshes enable traffic shifting and collaborative routing \cite{Concury_loadBalancer,collaborativeMesh_Infocom25,collaborativeMesh_TSC2025,istio,Envoy}.
Foundational load-balancer and transport systems such as Ananta, Maglev, and pFabric further motivate treating traffic engineering as an important determinant of microservice tail latency \cite{Ananta_SIGCOMM13,Maglev_NSDI16,pFabric_SIGCOMM13}.
Joint placement--routing optimization explicitly couples network dynamics with dependency structure \cite{OptTraffic_ICPP23,deployment_routing_TPDS23,TraDE_TPDS2026}. Production-trace characterization from Meta and Alibaba also suggests that communication modes (e.g., Inter-Process Communication (IPC), Remote Procedure Call (RPC), and Message Queue (MQ)) can materially affect microservice behavior and therefore deserve consideration when modeling dynamics \cite{github_metaTraces,github_AlibabaTraces,GrahGeneration_ICS25}.

\textbf{CI: Contention/interference dynamics.} CI captures noisy-neighbor and co-location effects driven by shared CPU, memory, cache, I/O, or network resources. Representative mechanisms include interference detection and prediction, resource shaping, and isolation in shared clusters
\cite{BubbleUp_colocation,Bubble_flux,PARTIES,CPI2_perfIsolation_cpu,PIMCloud_LCjobs,CLITE_colcation_LCjobs,Dirigent_LC_QoS},
as well as heterogeneity- and interference-aware cluster management \cite{Quasar_QoS_aware_clusterManagement,Paragon_QoS_hetero}.
Public-cloud measurement studies reveal how such interference appears as QoS degradation at scale \cite{Cloud_White},
and microservice-specific managers incorporate per-service latency sensitivity \cite{Erms_ASPLOS23,Derms_ISCA2024}.
A key observation is that interference often manifests primarily in tail latency, making SLO-aware control essential.

\textbf{FD: Failure/degradation dynamics.} FD captures failures, stragglers, availability loss, and QoS degradation events that require detection or mitigation. Cluster managers such as Borg and Fuxi incorporate recovery and blacklisting mechanisms \cite{Borg_clusterManager,Fuxi_alibaba}; diagnosis-oriented systems such as Sage, Seer, ART, PerfScope, and Orca localize degradation through telemetry \cite{Sage_ASPLOS21,Seer,ART_ASE24,PerfScope_bug_inference,Orca_bug_location}; and reliability-aware placement in fog and edge treats failures and partitions as the dominant dynamics \cite{ms_placement_iot_TMC,k8s_fog_latency,Net_ms_placement_TNSM}.

\textbf{SU: Sustainability/carbon dynamics.} SU captures time-varying energy budgets, power caps, renewable availability, and carbon-intensity signals that influence control decisions. Energy-aware autoscaling and VM allocation trade off performance and power \cite{saxena2022proactive},
and power management for latency-critical services illustrates how energy control interacts with tail latency \cite{Hipster_LC_task_manager,Rubik_LC_powerManagement}.
Geo-distributed workload management can exploit green-energy heterogeneity \cite{sCloud_goodput},
while recent survey work highlights carbon-aware resource management as an emerging direction for latency-sensitive clouds \cite{carbon_aware_yang2025survey}.
Recent web-service systems such as CASPER make this direction operational under explicit SLO constraints, while contemporaneous work on carbon-aware shifting cautions that achievable benefits depend strongly on workload flexibility and latency tolerance \cite{CASPER_IGSC23,CarbonShiftLimit_EuroSys24}.

\textbf{Takeaway of (D2).} Many systems model one or two dynamics explicitly. However, handling \emph{coupled} dynamics (e.g., service dependency + traffic imbalance + interference) is significantly harder due to partial observability and interacting controllers, which motivates the multi-dynamics focus of this survey.

\subsection{Dimension 3: Adaptation Strategy}
\label{subsec:D3}
D3 records how adaptation is realized. Adaptation strategies specify how a system \emph{senses}, \emph{decides}, and \emph{acts}. Many designs can be interpreted through a MAPE-K (Monitor, Analyze, Plan, Execute over Knowledge) or autonomic-computing lens \cite{autonomic_computing2003}, but they differ in the modeling formalism, optimization type, and available actuators. A work may be predictive and scaling-oriented at the same time because one category describes how a decision is produced, and the other describes how that decision is enacted. Thus, D3 explicitly separates \emph{decision logic} from \emph{actuation mechanism}.

\textbf{D3a: Decision logic.}

\textbf{RR: Reactive/rule-based.} RR denotes feedback rules, threshold policies, or direct control heuristics driven by current signals rather than an explicit predictive or learned model. Industrial autoscaling primitives such as HPA, VPA, the Cluster Autoscaler, KEDA, and Knative fit this category naturally \cite{K8s_HPA,k8s_vpa,k8s_cluster_autoscaler,keda,knative}, as do many runtime isolation or dataplane policies whose strength lies in simplicity and robustness rather than in long-horizon foresight.

\textbf{PM: Predictive/model-based.} PM denotes controllers or controller-support systems that use explicit forecasting, analytical models, or hybrid performance models to anticipate future behavior. Representative examples include workload forecasters \cite{CloudInsight_wokrloadPre,DeepTCN,esDNN_minxian,chen2023enbeats}, analytical performance prediction \cite{PREPARE_perfPredict,imporvedQueueNetworks}, and dependency-aware or application-aware autoscaling schemes such as pHPA, Madu, and Erlang \cite{Madu_SoCC22,pHPA_APNet21,Erlang_EuroSys24}. However, model accuracy and robustness under distribution shift directly determine whether proactive control is beneficial or harmful.

\textbf{OP: Optimization/planning.} OP denotes controllers that solve an explicit allocation, placement, scheduling, or routing problem at each decision epoch or planning horizon. This includes cluster scheduling and placement systems such as Borg, Fuxi, Quasar, Paragon, NetMARKS, and cloud--edge placement formulations \cite{Borg_clusterManager,Fuxi_alibaba,Quasar_QoS_aware_clusterManagement,Paragon_QoS_hetero,NetMARKS_InfoCOM21,adaptive_ms_cloudEdge_TPDS21,Adaptive_ms_IPDPS21,delay_edge_SPE22,delay_ms_TMC22}, as well as joint placement--routing designs \cite{OptTraffic_ICPP23,deployment_routing_TPDS23,TraDE_TPDS2026}.

\textbf{LH: Learning/hybrid.} LH denotes learning-based or explicitly hybrid controllers, especially reinforcement-learning designs that combine learned policies with heuristics, safety guards, or model-based components. Reinforcement learning (RL) and hybrid controllers (e.g., RL combined with heuristics or model-based components) are increasingly used to handle complex dynamics and multi-objective trade-offs (e.g., latency, cost, and utilization). Representative examples include $\mu$ConAdapter, CoScal, Safe RL for microservices, and hybrid energy-aware managers such as Hipster \cite{ms_rl_SOCC23,ms_rl_TNSM,SafeDRL_TC24,Hipster_LC_task_manager}. LH is useful for multi-objective control and complex state spaces, but it raises questions about sample efficiency, safety, and transferability.

\textbf{DG: Diagnosis-guided.} DG denotes telemetry-driven reasoning that localizes anomalies, performance regressions, or causal structure and then informs subsequent remediation. Sage, Seer, ART, PerfScope, and related AIOps-style systems exemplify this logic class \cite{Sage_ASPLOS21,Seer,ART_ASE24,PerfScope_bug_inference}. DG is important because some influential systems are adaptation-enabling rather than directly actuating.

\textbf{D3b: Actuation mechanism.}

\textbf{SC: Scaling.} SC adapts the number of replicas and/or the resources allocated to each replica.
Reactive autoscaling is common in practice (e.g., HPA and VPA) \cite{K8s_HPA,k8s_vpa}, while research explores proactive scaling driven by prediction and modeling \cite{CloudInsight_wokrloadPre,PREPARE_perfPredict,chen2023enbeats}.
Learning-based scaling and adaptation apply to both hard resources (e.g., per-service replicas and resources) and soft resources (e.g., concurrency of running threads or database connections) \cite{ms_rl_TNSM,ms_rl_SOCC23,SafeDRL_TC24}.
Cluster-level scaling with node provisioning interacts with pod scheduling and service-level scaling, often via the Cluster Autoscaler \cite{k8s_cluster_autoscaler}.
Recent microservice-specific scaling systems increasingly model service dependency structure and application-level objectives rather than scaling each service independently. Representative examples include SHOWAR for joint right-sizing and scheduling, Madu and DeepScaling for proactive or production-scale autoscaling, pHPA for microservice chains, and Erlang for application-aware cost and SLO trade-offs \cite{SHOWAR_SoCC21,Madu_SoCC22,DeepScaling_SoCC22,pHPA_APNet21,Erlang_EuroSys24}.
\emph{Practical implication:} scaling policies are prone to oscillations under multiple interacting controllers, which motivates dynamics-aware designs and implementations.

\textbf{PL: Placement/scheduling/migration.} PL denotes decisions about where the microservices run, taking into account system resource capacity, interference, reliability, and network costs. This includes Kubernetes scheduler extensions and network-aware scheduling \cite{k8s_network,NetMARKS_InfoCOM21,ExtendingNetK8s_ICSOC22,ExtendingNetK8s_CCGRID22,k8s_optim_ms}, reliability-aware service placement at fog/edge sites \cite{Net_ms_placement_TNSM,ms_placement_iot_TMC,Pallewatta_CSUR23},
cloud--edge deployment and migration \cite{adaptive_ms_cloudEdge_TPDS21,Adaptive_ms_IPDPS21,delay_edge_SPE22,delay_ms_TMC22}, and cost-efficient deployment formulations \cite{cost_efficient_MS}.
\emph{Practical implication:} service placement changes can also reshape call graphs and traffic patterns, creating feedback signals that cannot be ignored.

\textbf{RT: Routing/traffic management.} RT denotes traffic shifting, request steering, load balancing, retry policies, and collaborative routing, which are often enabled by service meshes and load balancers. Research explores traffic-aware optimization for replicated microservices \cite{OptTraffic_ICPP23}, load-balancing and routing mechanisms under traffic awareness \cite{Concury_loadBalancer,collaborativeMesh_Infocom25,collaborativeMesh_TSC2025}, and joint placement and routing under network and call-graph dynamics \cite{deployment_routing_TPDS23,TraDE_TPDS2026}. Production service-mesh stacks (e.g., Istio/Envoy) provide the operational primitives for such traffic control \cite{istio,Envoy}. Recent service-mesh work also studies architectural alternatives and verification-oriented tooling, including sidecar-free meshes and end-to-end policy testing \cite{CanalMesh_SIGCOMM24,ServiceMeshPolicies_ICPE22,MeshTest_NSDI25}. \emph{Practical implication:} traffic shifting can mitigate hotspots quickly, but it can also amplify control-loop oscillations when combined naively with autoscaling and rescheduling.

\textbf{IR: Isolation/resource shaping.} IR denotes mechanisms that partition, reserve, or shape shared resources to mitigate contention and tail-latency amplification. Mechanisms such as CPU isolation, admission control, and bandwidth allocation can mitigate resource interference. Representative approaches include co-location management and CPU-isolation mechanisms such as Heracles, FIRM, ERMS, DERM, PARTIES, Bubble-Up/Bubble-Flux, and MemGuard \cite{Heracles,FIRM_MS,Erms_ASPLOS23,Derms_ISCA2024,PARTIES,BubbleUp_colocation,Bubble_flux,MemGuard_mem_isolate}, as well as datacenter-network control for bandwidth isolation and deadline-aware transport \cite{D2TCP_netBand_isolate,Fastpass_netBand_isolate,EyeQ_isolate_netBand,netBand_isolate}. \emph{Practical implication:} isolation often acts as a ``safety layer'' that limits tail-latency spikes when predictions or policies are wrong.

\textbf{RE: Remediation/reconfiguration.} RE denotes restart, failover, rollback, blacklist, or other post-diagnosis corrective reconfiguration actions. Systems leverage telemetry to detect anomalies, infer dependencies, and guide remediation actions. Metrics tooling (e.g., Prometheus) and service-mesh telemetry (e.g., Istio) support online monitoring and anomaly detection \cite{Prometheus,NetMARKS_InfoCOM21,CoNEXT21_colocationMS},
while tracing frameworks (e.g., Jaeger) enable request-path and critical-path latency analysis \cite{dapper_tracing2010,jaeger,zipkin,opentelemetry}.
Representative remediation and AIOps-style systems leverage this telemetry to localize incidents and performance bottlenecks \cite{Sage_ASPLOS21,Seer,ART_ASE24,PerfScope_bug_inference}. In Table~\ref{tab:taxonomy-d1d4}, D3b is left as \texttt{--} when a representative work is primarily predictive, diagnostic, or empirical and does not expose a concrete actuator clearly enough in the current manuscript-level coding. \emph{Practical implication:} converting telemetry into actionable, \emph{causal} control signals remains difficult under sampling, noise, and evolving call graphs.

\textbf{Takeaway of (D3).}
D3 captures \emph{how} the adaptation strategy is realized in terms of decision logic and actuation mechanism. This clarifies, for example, why two systems can both be ``scaling'' yet differ fundamentally in whether they are reactive, predictive, optimized, learned, or diagnosis-guided.

\subsection{Dimension 4: Evaluation Evidence}
\label{subsec:D4}
D4 records the evidence basis of a system's claims. The choice of evaluation methodology significantly influences conclusions in dynamics-aware management, as realism, controllability, and reproducibility often involve trade-offs. 

\textbf{D4a: Execution substrate (PD/RC/EM/SM/OF).} D4a records the substrate on which the controller or study runs: production-deployed systems (PD), real-cluster or real-host deployments (RC), emulation (EM), simulation (SM), or offline trace-, dataset-, or model-only study (OF). Real clusters and production deployments remain the strongest evidence modes for end-to-end controller validation, but they are often harder to reproduce \cite{github_AlibabaTraces,github_metaTraces,metaTraces_ATC23,Borg_trace,Alibaba_colocatedData}. Emulation provides better control over network and infrastructure perturbations \cite{Kollaps_EuroSys20,THUNDERSTORM_SRDS19}, while simulation supports larger what-if studies and design-space exploration \cite{CloudSim_SPE2011,Cloudnativesim_SPE25}. Offline studies include workload characterization, predictive modeling, and trace/dataset analysis where no live managed system or controller is executed \cite{chen2023enbeats,DeepTCN,esDNN_minxian,microsoftResCentral_SOSP17,SoCC2021_MS_luo}.

\textbf{D4b: Workload source (SY/BM/TD).} D4b records whether the dynamic signal driving evaluation comes primarily from synthetic workload scenarios (SY), benchmark suites (BM), or trace-driven studies (TD). Benchmark suites such as DeathStarBench and $\mu$Bench improve repeatability for microservice evaluations \cite{deathStarBench_ASPLOS19,muBench_TPDS23}; TrainTicket supports controlled debugging and failure-oriented studies \cite{trainTicket}; and production traces from Alibaba and Meta capture non-stationary demand and evolving dependencies that synthetic tests often miss \cite{SoCC2021_MS_luo,metaTraces_ATC23,ACMDatasetsCloud2025}. D4 separates these workload-source choices from the execution substrate, as a benchmark can be run either on a real cluster or in a simulation tool.

\textbf{D4c: Fidelity enhancer (KL/NE/FI/NO).} D4c records whether the evidence is strengthened by an explicit fidelity enhancer. We use KL when Kubernetes control-plane behavior, Kubernetes APIs, or Kubernetes-native actuation are part of the evaluated management loop, not merely when a benchmark is deployed on a Kubernetes cluster \cite{k8sLoop_simu_PerformanceTools2024,chen2025idynamics}. We use NE for network emulation or explicit network shaping \cite{Kollaps_EuroSys20,THUNDERSTORM_SRDS19}, FI for fault injection or chaos-style perturbation \cite{ChaosEngineering_IEEE16,FailureTesting_SoCC16,NetflixFailure_LISA13}, and NO when no such fidelity enhancer is made explicit in the proposed approach. This sub-dimension is important for cross-layer controllers because simplified orchestration or communication models can hide exactly the interactions that later dominate deployment behavior.


\textbf{Takeaway of (D4).}
D4 strongly conditions how confidently conclusions can be interpreted: performance gains observed in simplified evaluation settings may disappear once realistic orchestration, network dynamics, and interference are incorporated.

\begin{table*}[!t]
\centering
\caption{Comparative synthesis across major controller families in the coded corpus of representative systems. Instead of a tag-only summary, this table makes explicit the state assumptions, operational timescales, telemetry burden, coupling structure, typical evidence base, and recurring transfer risks for each family in our coded corpus. Timescales are qualitative (fast path, control epoch, or slower planning loop) rather than implementation-specific clock settings.}
\label{tab:comparative-synthesis}

{\tiny
\setlength{\tabcolsep}{1.6pt}
\renewcommand{\arraystretch}{0.96}
\setlength{\arrayrulewidth}{0.25pt}

\newcommand{\fulltableline}{\hhline{--|---|--}}
\newcommand{\grouptableline}{\hhline{~~|---|--}}

\begin{tabular}{@{}>{\raggedright\arraybackslash}p{0.10\linewidth}
                    >{\raggedright\arraybackslash}p{0.14\linewidth}|
                    >{\raggedright\arraybackslash}p{0.10\linewidth}
                    >{\raggedright\arraybackslash}p{0.12\linewidth}
                    >{\raggedright\arraybackslash}p{0.13\linewidth}|
                    >{\raggedright\arraybackslash}p{0.20\linewidth}
                    >{\raggedright\arraybackslash}p{0.14\linewidth}@{}}

\fulltableline
\multirow{2}{=}{\textbf{Family}} &
\multirow{2}{=}{\textbf{State assumptions}} &
\multicolumn{3}{|c|}{\textbf{Control characteristics}} &
\multicolumn{2}{c}{\textbf{Evidence and Risk}} \\
\grouptableline
& &
\textbf{Timescale} &
\textbf{Telemetry} &
\textbf{Coupling structure} &
\textbf{Evidence base} &
\textbf{Transfer risk} \\
\fulltableline

Autoscaling \& right-sizing (\cite{Madu_SoCC22,DeepScaling_SoCC22,pHPA_APNet21,Erlang_EuroSys24}) &
Per-service demand and resource state can be inferred from short-horizon signals; downstream or cross-service effects are either ignored or compressed into coarse summaries. &
Control epoch (seconds to minutes) &
CPU and memory utilization, request rate, queue length or backlog, and SLO metrics &
Usually workload--resource coupling; dependency coupling appears in chain-aware or graph-aware variants &
Forecasting support often uses trace-driven offline or simulation evidence, while controller papers are mainly evaluated with benchmark-driven real-system testbeds; DeepScaling provides production-deployed evidence in this family &
Misprediction, lagged actuation, or cooldown interactions can shift bottlenecks downstream and induce oscillations \\
\fulltableline

Graph-aware reasoning \& incident control (\cite{Parslo_SoCC21,ChainFormer_ICSOC23,Sage_ASPLOS21,Seer,SoCC2021_MS_luo}) &
The dependency graph and critical path can be recovered from traces with sufficient stability for localization or control. &
Diagnosis or control epoch (seconds to minutes) &
Distributed traces and service-level metrics &
Call-graph coupling is explicit; network, failure, and interference effects are handled unevenly across systems &
Mixed evidence: real-system testbeds for control or localization systems, benchmark and trace-driven workloads for graph-aware controllers, and offline trace characterization when the work is primarily an evidence source rather than a controller &
Trace sampling, graph drift, and partial observability weaken attribution, root-cause localization, and SLO budgeting \\
\fulltableline

Placement, routing \& service-mesh control (\cite{NetMARKS_InfoCOM21,k8s_network,TraDE_TPDS2026,Concury_loadBalancer,collaborativeMesh_TSC2025}) &
Topology, locality, and network state are observable at placement or routing time with adequate freshness for control. &
Fast routing path; slower placement or reconfiguration loop &
Scheduler state, service-mesh counters, latency and bandwidth measurements, and topology signals &
Placement--network coupling is explicit; some systems also model call-graph or edge-placement coupling &
Mostly real-system testbeds for Kubernetes or service-mesh control; simulation remains common in edge/fog and reliability studies, and explicit emulation or network shaping appears in a small number of network-focused systems &
Stale network views, orchestration delay, or policy interactions can erode gains and create routing or placement churn \\
\fulltableline

Interference-aware isolation \& resource control (\cite{FIRM_MS,Erms_ASPLOS23,Derms_ISCA2024,PARTIES,BubbleUp_colocation,Bubble_flux}) &
Contention signatures can be detected reliably enough to trigger isolation or resource shaping. &
Fast safety loop with slower reallocation &
Resource metrics, hardware counters, service latency, and tail-latency metrics &
Co-location coupling is explicit; end-to-end dependency coupling is often only implicit or handled through critical-path summaries &
Predominantly real-system testbeds with benchmark or synthetic workloads; several systems add trace-driven simulation or fault/contention injection, but explicit platform-in-the-loop fidelity remains sparse &
Local QoS protection may miss upstream or downstream bottlenecks or side effects across interacting control loops \\
\fulltableline

Learning-based multi-knob control (\cite{ms_rl_SOCC23,ms_rl_TNSM,SafeDRL_TC24}) &
The training environment, reward proxy, and state representation sufficiently approximate deployment conditions. &
Controller epoch for actuation; learning horizon spans many episodes &
Aggregated metrics and SLO signals encoded as state vectors &
Potentially multi-knob, but the learned coupling is often opaque and difficult to interpret &
Mixed evidence: real-system prototypes for microservice RL controllers and trace-driven simulation for edge reliability/provisioning; no production-deployed evidence appears in the coded learning-specific rows &
Distribution shift, unsafe exploration, and hidden actuation costs reduce trustworthiness and transferability \\
\fulltableline

Sustainability-aware control (\cite{saxena2022proactive,Hipster_LC_task_manager,Rubik_LC_powerManagement,CarbonAware_2023}) &
Energy or carbon signals are available at useful granularity, and the workload has enough temporal or spatial flexibility to exploit them. &
Slower planning loop (minutes to hours) &
Power, energy, or carbon signals, together with utilization and SLO metrics &
Latency--cost--energy trade-offs are explicit; call-graph and network coupling are usually weak &
Evidence spans production carbon-aware capacity management, trace-driven simulation, and real-system energy/tail-latency testbeds; microservice-specific production evidence remains limited &
Benefits depend strongly on latency slack, workload shiftability, and the fidelity of energy or carbon signals \\
\fulltableline

\end{tabular}
}
\end{table*}

\subsection{Orthogonal Dimensions: Objectives and Telemetry}
\label{subsec:crosscutting}
Although not treated as primary dimensions in our classification table, objectives and telemetry repeatedly shape the design of dynamics-aware systems.

\textbf{Objectives and SLO models.}
Much of microservice management work is SLO-driven, with tail latency and throughput as dominant objectives \cite{tail_at_scale2013,deathStarBench_ASPLOS19}.
Cloud systems such as FIRM \cite{FIRM_MS}, ERMS \cite{Erms_ASPLOS23}, and DERM \cite{Derms_ISCA2024} target SLO/SLA assurance while improving utilization,
and cost-aware deployment and consolidation introduce explicit cost--performance trade-offs \cite{cost_efficient_MS,sharedMS_AcmTranCS}.
Sustainability and energy objectives are also receiving increasing attention, including energy-efficient autoscaling and power management \cite{saxena2022proactive,Hipster_LC_task_manager,Rubik_LC_powerManagement},
as well as emerging directions in carbon-aware resource management \cite{carbon_aware_yang2025survey}.
A recurring concern is that objectives interact through trade-offs. For instance, minimizing latency can increase cost or energy consumption, while aggressive consolidation can exacerbate interference and tail latency.

\textbf{Telemetry signals.}
The available telemetry constrains which dynamics can be detected and controlled.
Metrics systems such as Prometheus provide time-series signals used in autoscaling and anomaly detection \cite{Prometheus},
and service-mesh telemetry can expose fine-grained network interaction signals useful for scheduling and co-location \cite{NetMARKS_InfoCOM21,CoNEXT21_colocationMS}.
Tracing frameworks (e.g., Jaeger) reveal request paths and critical-path latency breakdowns \cite{dapper_tracing2010,jaeger,zipkin,opentelemetry},
thereby enabling call-graph-aware debugging and dependency-aware control \cite{Sage_ASPLOS21,Parslo_SoCC21,ChainFormer_ICSOC23,Seer}.
Classic and recent tracing work further shows that telemetry quality depends on preserving causality and representative rare paths under sampling and overhead constraints \cite{XTrace_NSDI07,PivotTracing_SOSP15,TProf_SoCC21,STEAM_FSE23,DeepFlow_SIGCOMM23}.
However, telemetry is noisy, sampled, and incomplete, meaning that converting it into robust causal signals remains a core challenge \cite{Sage_ASPLOS21,Seer,PerfScope_bug_inference}.

To make the cross-literature comparison explicit, rather than leaving it as a coding-only inventory, Table~\ref{tab:comparative-synthesis} synthesizes recurring patterns across the major controller families in the coded corpus. Compared with a tag-only mapping, it highlights implicit state assumptions, control horizons, telemetry burden, coupling structure, and generalization risks that often determine whether a method transfers beyond its original testbed. We retain the detailed per-system coding in Table~\ref{tab:taxonomy-d1d4} so that readers can trace these comparative judgments back to representative systems.

{
\tiny   
\setlength{\tabcolsep}{1.6pt}
\renewcommand{\arraystretch}{1.2}
\begin{longtable}{>{\raggedright\arraybackslash}p{0.13\linewidth} >{\centering\arraybackslash}p{0.08\linewidth} >{\centering\arraybackslash}p{0.06\linewidth} >{\centering\arraybackslash}p{0.06\linewidth} >{\centering\arraybackslash}p{0.06\linewidth} >{\centering\arraybackslash}p{0.06\linewidth} >{\centering\arraybackslash}p{0.06\linewidth} >{\centering\arraybackslash}p{0.06\linewidth} >{\raggedright\arraybackslash}p{0.40\linewidth}}
\caption{Detailed coded corpus of representative systems mapped to taxonomy dimensions D1--D4. Pure evaluation artifacts (benchmarks, traces, simulators, emulators, and in-the-loop tools) are consolidated in Table~\ref{tab:benchmarks-traces} to avoid duplication. This coded corpus underpins the comparative synthesis in Table~\ref{tab:comparative-synthesis} and the descriptive evaluation observations in Section~\ref{subsec:eval-repro}.}\\
\label{tab:taxonomy-d1d4} \\
\hline
\textbf{Work} & \textbf{D1} & \textbf{D2} & \textbf{D3a} & \textbf{D3b} & \textbf{D4a} & \textbf{D4b} & \textbf{D4c} & \textbf{Key idea} \\
\hline
\endfirsthead
\multicolumn{9}{c}{\textbf{Table~\thetable{} continued.}} \\
\hline
\textbf{Work} & \textbf{D1} & \textbf{D2} & \textbf{D3a} & \textbf{D3b} & \textbf{D4a} & \textbf{D4b} & \textbf{D4c} & \textbf{Key idea} \\
\hline
\endhead
\hline
\multicolumn{9}{p{0.96\linewidth}}{\tiny
\textit{Legend.}
D1: HV = host/VM; CO = cluster-orchestrator; SV = service;
AG = application/request graph; DP = dataplane/service-mesh/network;
\texttt{[ED]} = explicit edge or multi-cluster deployment modifier.
D2: WS = workload/service demand; RD = request path/dependency;
NS = network state; CI = contention/interference; FD = failure/degradation;
SU = sustainability/carbon signal.
D3a: RR = reactive/rule-based; PM = predictive/model-based;
OP = optimization/planning; LH = learning/hybrid; DG = diagnosis-guided.
D3b: SC = scaling; PL = placement/scheduling/migration;
RT = routing/traffic management; IR = isolation/resource shaping;
RE = remediation/reconfiguration; \texttt{--} = no explicit actuation
surface in the representative paper.
D4a: PD = production-deployed; RC = real-cluster/host testbed;
EM = emulation; SM = simulation; OF = offline trace-, dataset-, or
model-only study.
D4b: SY = synthetic or load-generator workload; BM = benchmark-driven;
TD = trace-driven.
D4c: KL = Kubernetes-in-the-loop; NE = explicit network shaping or
emulation; FI = fault injection or chaos; NO = no explicit fidelity
enhancer. For platform/documentation entries, \texttt{N/A} indicates that a coding
dimension is not applicable to an evaluated research artifact.
}
\endlastfoot

\rowcolor{blue!10}\multicolumn{9}{l}{\textit{Predictive modeling and proactive scaling}} \\
\hline
EN-Beats \cite{chen2023enbeats} & HV/CO & WS & PM & -- & OF & TD & NO & Offline multi-resource forecasting on real cloud traces; useful for proactive control but no evaluated actuator. \\
\hline
CloudInsight \cite{CloudInsight_wokrloadPre} & HV/CO & WS & PM & SC & SM/OF & TD & NO & Ensemble workload forecasting with trace-based resource-management simulation for predictive scaling. \\
\hline
DeepTCN \cite{DeepTCN} & -- & WS & PM & -- & OF & TD & NO & Generic probabilistic time-series forecasting baseline on real-world traces; background-only unless tied explicitly to cloud/microservice control. \\
\hline
esDNN \cite{esDNN_minxian} & HV/CO & WS & PM & SC & SM/OF & TD & NO & Multivariate cloud workload prediction with simulated machine/resource scaling on Alibaba/Google traces. \\
\hline
Resource Central \cite{microsoftResCentral_SOSP17} & CO/HV & WS, CI & PM & PL/IR & SM/OF & TD & NO & Azure VM telemetry and prediction service; scheduler changes are evaluated mainly with real VM traces, while models/features were used in production for analysis. \\
\hline
PREPARE \cite{PREPARE_perfPredict} & HV & WS, FD & PM/DG & RE & RC & BM, TD & FI & Predicts performance anomalies and triggers corrective actions such as VM migration; evaluated on RUBiS/System S plus traces with injected faults. \\
\hline
\cite{Arima_ref3} & SV/CO & WS & PM/OP & SC & SM & TD & NO & Prediction-driven autoscaling using receding-horizon and MVA-style planning over workload traces. \\
\hline

\rowcolor{teal!10}\multicolumn{9}{l}{\textit{Industry autoscaling and orchestration primitives}} \\
\hline
K8s HPA \cite{K8s_HPA} & SV/CO & WS & RR & SC & PD & N/A & NO & Production Kubernetes primitive for metric-triggered horizontal pod autoscaling. \\
\hline
K8s VPA \cite{k8s_vpa} & SV/CO & WS & RR & SC & PD & N/A & NO & Production Kubernetes primitive for vertical right-sizing of pod resource requests/limits. \\
\hline
K8s Autoscaler \cite{k8s_cluster_autoscaler} & CO & WS & RR & SC & PD & N/A & NO & Production cluster/node autoscaler reacting to unschedulable pods and utilization. \\
\hline
Karpenter \cite{karpenter} & CO & WS & OP & PL, SC & PD & N/A & NO & Production node-provisioning primitive with instance-type/capacity selection. \\
\hline
KEDA \cite{keda} & SV/CO & WS & RR & SC & PD & N/A & NO & Production event-driven autoscaling primitive for queue/stream/external triggers. \\
\hline
Knative \cite{knative} & SV & WS & RR & SC & PD & N/A & NO & Production serving primitive for request-driven scaling, scale-to-zero, and bursts. \\
\hline

\rowcolor{teal!28}\multicolumn{9}{l}{\textit{Microservice autoscaling and right-sizing}} \\
\hline
SHOWAR \cite{SHOWAR_SoCC21} & SV/CO & WS & OP/RR & SC, PL & RC & BM & NO & Kubernetes right-sizing plus scheduling hints for interactive microservice benchmarks. \\
\hline
Madu \cite{Madu_SoCC22} & SV/AG & WS, RD & PM & SC & RC & BM, TD & NO & Workload-learning autoscaler for multiplexed service dependencies; validated on private cluster with benchmark apps and Alibaba-derived traces. \\
\hline
DeepScaling \cite{DeepScaling_SoCC22} & SV/AG & WS, RD & LH & SC & PD & TD & NO & Production Ant Group autoscaler combining forecasting and DQN-style scaling for large microservice fleets. \\
\hline
pHPA \cite{pHPA_APNet21} & AG/SV & WS, RD & PM/OP & SC & RC & BM, SY & NO & Proactive chain autoscaling in Kubernetes using dependency-aware prediction/optimization and benchmark workloads. \\
\hline
Erlang \cite{Erlang_EuroSys24} & AG/SV/CO & WS, RD & OP/LH & SC & RC & BM, SY & NO & Application-aware autoscaling on GKE using profiling/optimization across open-source microservice applications. \\
\hline

\rowcolor{green!12}\multicolumn{9}{l}{\textit{Graph-aware and dependency-aware management}} \\
\hline
Sage \cite{Sage_ASPLOS21} & AG/SV & RD, FD, CI & DG/LH & RE & RC & BM, SY & FI, NE & ML-driven root-cause localization and corrective action under injected resource and network perturbations. \\
\hline
Parslo \cite{Parslo_SoCC21} & AG & RD, WS & OP & -- & RC/SM & BM, SY & NO & Dependency-aware partial SLO allocation for microservice DAGs; supports downstream resource controllers but does not itself expose a concrete actuator. \\
\hline
ChainsFormer \cite{ChainFormer_ICSOC23} & AG/SV & RD, WS & LH & SC & RC & BM, TD & NO & Chain-latency-aware RL resource provisioning on Kubernetes with benchmark services and Alibaba traces. \\
\hline
\cite{SoCC2021_MS_luo} & AG & RD, WS, CI & -- & -- & OF & TD & NO & Empirical Alibaba trace characterization of dependencies and performance; evidence source, not an adaptive controller. \\
\hline

\rowcolor{cyan!12}\multicolumn{9}{l}{\textit{Placement, scheduling, and edge/network-aware management}} \\
\hline
Borg \cite{Borg_clusterManager} & CO & WS, FD & OP/RR & PL, RE & PD & TD & NO & Production cluster manager with admission control, packing, scheduling, and failure recovery. \\
\hline
Fuxi \cite{Fuxi_alibaba} & CO & WS, FD & OP/RR & PL, RE & PD & SY, TD & FI & Alibaba production-scale resource manager/scheduler with fault tolerance. \\
\hline
Mercury \cite{Mercury_scheduling_throughput} & CO & WS & OP & PL & RC & BM, TD & NO & Hybrid centralized/distributed scheduler on a large real cluster using GridMix and trace-based jobs. \\
\hline
Quasar \cite{Quasar_QoS_aware_clusterManagement} & CO & WS, CI & PM/OP & PL, IR & RC & BM, SY & NO & QoS-aware cluster manager using classification for placement/resource allocation under heterogeneity and interference. \\
\hline
Paragon \cite{Paragon_QoS_hetero} & CO & CI, WS & PM/OP & PL & RC & BM, SY & NO & Interference- and heterogeneity-aware scheduler using lightweight profiling/classification. \\
\hline
NetMARKS \cite{NetMARKS_InfoCOM21} & CO/DP & NS, RD & OP & PL & RC & SY & NO & Service mesh network-metric-aware Kubernetes scheduler for latency/locality-sensitive placement. \\
\hline
Diktyo (K8s) \cite{k8s_network} & CO & NS & OP & PL & RC & BM, SY & NO & Network-aware Kubernetes scheduler using measured communication costs and application workloads. \\
\hline
\cite{ExtendingNetK8s_ICSOC22} & CO[ED] & NS & OP & PL & RC & SY & NO & Network-aware Kubernetes scheduling extensions for cloud-to-edge placement. \\
\hline
\cite{k8s_interference_TCC} & CO & CI, WS & OP & PL, IR & SM & SY & NO & Kubernetes-oriented placement model with dynamic resource allocation under interference; mainly synthetic/algorithmic evaluation. \\
\hline
\cite{k8s_optim_ms} & CO & WS, NS & OP & PL & RC & BM, SY & NO & GCP/Kubernetes cost-aware microservice placement with real use cases and generated workloads. \\
\hline
\cite{cost_efficient_MS} & CO & WS, NS & OP & PL & SM/OF & TD & NO & Trace-driven layered-container placement and image-pulling optimization using registry traces. \\
\hline
OptTraffic \cite{OptTraffic_ICPP23} & CO/DP & NS, RD & OP & PL, RT & RC & BM, SY & NO & Kubernetes traffic-locality optimization for multi-replica microservices. \\
\hline
TraDE \cite{TraDE_TPDS2026} & AG/CO/DP & RD, NS & OP & PL, RT & RC & BM, SY & NE, KL & Kubernetes extension for joint deployment and routing under dynamic cross-node delay conditions. \\
\hline
\cite{Net_ms_placement_TNSM} & CO[ED] & NS, FD & OP & PL & SM & SY & NO & Simulation-based network-aware reliability modeling and placement for microservice in fog environments. \\
\hline
\cite{ms_placement_iot_TMC} & CO[ED] & NS, FD & OP & PL & SM & SY, TD & NO & Reliability-aware proactive placement for IoT/fog microservices using iFogSim-style simulation and derived failure traces. \\
\hline
\cite{adaptive_ms_cloudEdge_TPDS21} & CO/SV[ED] & NS, CI, WS & LH/OP & PL, IR & RC & BM & NO & Cloud-edge microservice deployment and resource allocation considering communication and contention. \\
\hline
\cite{Adaptive_ms_IPDPS21} & CO/SV[ED] & NS, CI, WS & LH/OP & PL, IR & RC & BM & NO & QoS-aware cloud-edge deployment with joint placement and resource allocation on a real cloud-edge continuum. \\
\hline
PDMA \cite{delay_edge_SPE22} & CO[ED] & NS, WS & OP/PM & PL & SM & TD & NO & Delay- and mobility-aware service migration in MEC using simulation driven by different workload datasets. \\
\hline
\cite{delay_ms_TMC22} & CO[ED] & NS, WS & LH & PL & SM & TD & NO & RL-based microservice coordination and migration in MEC using trace-driven Monte Carlo simulation. \\
\hline
\cite{k8s_fog_latency} & CO[ED] & NS, FD & OP & PL, RE & RC & SY & FI & Kubernetes-based fog orchestration with latency-aware allocation and failover experiments. \\

\hline
\rowcolor{violet!12}\multicolumn{9}{l}{\textit{Traffic management and service meshes}} \\
\hline
Concury \cite{Concury_loadBalancer} & DP & NS, WS & OP & RT & RC/EM & SY, TD & NE & Software load-balancer evaluated on DPDK/CloudLab-style networks and Mininet/P4 with real and synthetic traffic. \\
\hline
\cite{collaborativeMesh_Infocom25} & DP & RD, NS, WS & OP & RT & RC & BM, SY & NO & Probabilistic collaborative service-mesh routing/orchestration in a private Istio cluster. \\
\hline
\cite{collaborativeMesh_TSC2025} & DP & RD, NS, WS & OP & RT & RC/SM & BM, SY & NO & Large-scale service-mesh orchestration with probabilistic routing over benchmark applications and simulations. \\
\hline
Canal Mesh \cite{CanalMesh_SIGCOMM24} & DP & NS, CI & OP/RR & RT, IR & PD/RC & TD, SY & NO & Alibaba sidecar-free service-mesh architecture reducing dataplane/control-plane overhead, with production data and testbed evaluation. \\
\hline
Istio \cite{istio} & DP & NS, FD & RR & RT, RE & PD & N/A & NO & Production service-mesh primitive for traffic shifting, retries, policy, and failure handling. \\
\hline
Envoy \cite{Envoy} & DP & NS, FD & RR & RT, RE & PD & N/A & NO & Production proxy/dataplane primitive for routing, resilience, and traffic policy enforcement. \\
\hline
Linkerd \cite{linkerd} & DP & NS, FD & RR & RT, RE & PD & N/A & NO & Production service-mesh primitive for routing, observability, and reliability features. \\

\hline
\rowcolor{orange!14}\multicolumn{9}{l}{\textit{Interference-aware and SLO/SLA-aware resource management}} \\
\hline
Heracles \cite{Heracles} & HV/SV & WS, CI & RR/OP & IR & RC/PD & TD, SY & NO & Co-location manager for latency-critical services using production workloads and interference characterization. \\
\hline
FIRM \cite{FIRM_MS} & AG/SV/HV & RD, CI, WS & LH/DG & IR, SC & RC & BM, SY & FI, NE & Kubernetes microservice resource manager with ML/RL, critical-path diagnosis, and injected resource and network anomalies. \\
\hline
Erms \cite{Erms_ASPLOS23} & AG/SV & RD, WS, CI & OP & IR, SC & RC/SM & BM, TD & FI & Shared-microservice SLA management with real deployments, trace-driven simulations, and injected interference. \\
\hline
Sinan \cite{Sinan_MS} & AG/SV & RD, WS, CI & PM/LH & IR & RC & BM, SY & NO & ML-driven online microservice resource manager for DeathStarBench applications under dynamic load. \\
\hline
GrandSLAm \cite{grandslam_MS} & AG/SV & RD, WS & OP/RR & IR, RT & RC & BM, SY & NO & SLA-aware request and resource scheduling for multi-stage microservice execution frameworks. \\
\hline
Derm \cite{Derms_ISCA2024} & AG/SV/HV & RD, WS, CI & OP & IR, SC & RC/SM & BM, TD & NO & SLA-aware resource management for highly dynamic microservice graphs using benchmarks and trace-driven simulation. \\
\hline
\cite{sharedMS_AcmTranCS} & AG/SV & RD, WS, CI & OP & IR, SC & RC/SM & BM, TD & FI & Scalable shared-microservice resource management with benchmarks, Alibaba traces, and injected interference. \\
\hline
CLITE \cite{CLITE_colcation_LCjobs} & HV/SV & CI & OP/RR & IR & RC & BM, SY & NO & QoS-aware co-location of multiple latency-critical jobs using resource adjustment. \\
\hline
PIMCloud \cite{PIMCloud_LCjobs} & HV & CI, SU & OP/RR & IR & SM & BM, SY & NO & Cycle-level simulation of QoS-/energy-aware resource management for latency-critical workloads on PIM systems. \\
\hline
PARTIES \cite{PARTIES} & HV & CI & RR/OP & IR & RC & BM, SY & NO & QoS-aware resource partitioning for co-located interactive services. \\
\hline
Bubble-Up \cite{BubbleUp_colocation} & HV & CI & PM & PL/IR & RC/PD & TD, SY & NO & Bubble-based interference profiling to choose safe co-locations in warehouse-scale settings. \\

\hline
Bubble-Flux \cite{Bubble_flux} & HV & CI & PM/RR & IR & RC & BM, SY & NO & Online QoS management using interference prediction and feedback for co-located services. \\
\hline
CPI$^2$ \cite{CPI2_perfIsolation_cpu} & HV & CI & RR & IR & RC/PD & TD, SY & NO & CPU performance isolation using CPI-based interference detection in shared clusters. \\
\hline
PerfIso \cite{Perfiso_resAlo} & HV & CI & RR/OP & IR & RC/PD & BM, SY & NO & Performance isolation for commercial latency-sensitive services through resource allocation. \\
\hline
MemGuard \cite{MemGuard_mem_isolate} & HV & CI & RR & IR & RC & BM, SY & NO & Memory-bandwidth reservation to reduce multicore interference. \\
\hline
\cite{perfInterf_multicore} & HV & CI & PM & -- & RC & BM, SY & NO & Regression-based cross-core interference prediction; useful for consolidation decisions but no direct management actuator. \\
\hline
DeepDive \cite{DeepDive} & HV & CI & DG & IR/RE & RC & BM, SY & FI & Identifies and mitigates VM performance interference in virtualized environments. \\
\hline
Dirigent \cite{Dirigent_LC_QoS} & HV/CO & WS, CI & RR/OP & IR & RC & BM, SY & NO & QoS enforcement for latency-critical tasks on shared multicore systems. \\
\hline
\cite{CoNEXT21_colocationMS} & CO/DP & CI, NS & OP & PL & RC & BM, SY & NO & Service-mesh telemetry guided co-location for containerized workloads. \\
\hline

\rowcolor{red!10}\multicolumn{9}{l}{\textit{Network isolation and transport}} \\
\hline
D$^2$TCP \cite{D2TCP_netBand_isolate} & DP & NS & OP/RR & RT, IR & RC/SM & SY & NO & Deadline-aware datacenter transport allocating bandwidth to time-sensitive flows. \\
\hline
Fastpass \cite{Fastpass_netBand_isolate} & DP & NS & OP & RT, IR & RC & TD, SY & NO & Centralized datacenter network scheduling for near-zero queuing. \\
\hline
EyeQ \cite{EyeQ_isolate_netBand} & DP & CI, NS & RR/OP & IR & RC/EM & BM, SY & NO & Network performance isolation for multi-tenant cloud and edge settings. \\
\hline
PDQ \cite{netBand_isolate} & DP & NS & OP & RT, IR & SM & SY, TD & NO & Packet and flow-level simulation of preemptive datacenter flow scheduling. \\
\hline

\rowcolor{lime!14}\multicolumn{9}{l}{\textit{Sustainability-aware management}} \\
\hline
\cite{CarbonAware_2023} & CO[ED] & SU, WS & PM/OP & PL, IR & PD & TD & NO & Google production carbon-aware compute management via capacity shaping and flexible workload scheduling. \\
\hline
\cite{saxena2022proactive} & HV/CO & WS, SU & PM/OP & SC, PL & SM & TD & NO & Trace-driven simulation of proactive autoscaling and energy-efficient VM allocation. \\
\hline
Rubik \cite{Rubik_LC_powerManagement} & HV/SV & SU, CI & OP/RR & IR & RC & BM, SY & NO & Analytical power management for latency-critical systems under tail-latency constraints. \\
\hline
Hipster \cite{Hipster_LC_task_manager} & HV/SV & WS, SU, CI & LH/OP & IR & RC & BM, SY & NO & Hybrid task and core management for latency-critical workloads with energy awareness. \\
\hline
sCloud \cite{sCloud_goodput} & CO[ED] & SU, WS & OP & PL & SM & TD & NO & Trace-driven management of workloads across distributed sustainable datacenters. \\

\hline
\rowcolor{magenta!10}\multicolumn{9}{l}{\textit{Learning-based controllers}} \\
\hline
$\mu$ConAdapter \cite{ms_rl_SOCC23} & SV & WS, CI & LH & IR & RC & BM, TD & NO & RL-based concurrency adaptation for SockShop and SocialNetwork benchmarks under bursty traces. \\
\hline
CoScal \cite{ms_rl_TNSM} & SV/CO & WS & LH & SC, RE & RC & BM, TD & NO & RL-based multifaceted microservice adaptation combining horizontal and vertical scaling and brownout method. \\
\hline
SafeDRL \cite{SafeDRL_TC24} & SV/CO[ED] & WS, FD, NS & LH & PL, RE & SM & TD & NO & Safe DRL provisioning of primary and backup service functions in edge environments using trace-driven simulations. \\
\hline

\rowcolor{purple!15}\multicolumn{9}{l}{\textit{Diagnosis and remediation}} \\
\hline
Seer \cite{Seer} & AG/SV & RD, FD, CI & DG & -- & RC/PD & BM, SY & NO & Online performance debugging and localization for cloud microservices; external cluster manager acts on recommendations. \\
\hline
ART \cite{ART_ASE24} & AG/SV & FD, RD & DG & -- & OF/RC & BM, TD & FI & Unified unsupervised incident management over datasets from benchmark and production-like microservice systems. \\
\hline
PerfScope \cite{PerfScope_bug_inference} & HV/SV & FD & DG & -- & RC & BM, TD & NO & Online performance bug inference evaluated on real bugs in server systems on a virtualized cloud testbed. \\
\hline
Orca \cite{Orca_bug_location} & SV/AG & FD & DG & -- & PD & TD & NO & Differential commit-level bug localization actively used by OCEs in large-scale services. \\\hline
\end{longtable}
}

\subsection{Comparative Evaluation and Reproducibility Guidance}
\label{subsec:eval-repro}

Evaluation is not a secondary issue in dynamics-aware microservice management; it materially affects how confidently a claimed improvement can be interpreted. The literature already reveals a persistent trade-off between realism and repeatability. Benchmark suites such as DeathStarBench and $\mu$Bench improve comparability by providing reusable microservice applications and configurable bottlenecks, while TrainTicket enables controlled failure and debugging studies in realistic service graphs \cite{deathStarBench_ASPLOS19,muBench_TPDS23,trainTicket}. Production traces from Alibaba and Meta reveal non-stationary demand, evolving dependencies, and co-location effects that synthetic workloads often miss \cite{SoCC2021_MS_luo,metaTraces_ATC23,ACMDatasetsCloud2025}. At the same time, network emulation and Kubernetes-in-the-loop approaches have shown that simplified evaluations can hide controller interactions that only become apparent once realistic routing, orchestration delays, and failure handling are introduced \cite{Kollaps_EuroSys20,THUNDERSTORM_SRDS19,k8sLoop_simu_PerformanceTools2024,chen2025idynamics}. For failure-oriented evaluation, chaos-engineering and lineage-driven fault-injection studies show that perturbations should be systematic and dependency-aware rather than ad hoc \cite{ChaosEngineering_IEEE16,FailureTesting_SoCC16,NetflixFailure_LISA13}. For this reason, comparative evaluation should be organized around the \emph{dynamics being claimed}, rather than around a single convenient workload or testbed.

Beyond this normative checklist, the coded corpus reveals several concrete evaluation patterns. Table~\ref{tab:eval-patterns} summarizes descriptive patterns across the 75 research-paper entries defined in Section~\ref{sec:scope-corpus}, excluding the 9 platform/documentation entries. One clear pattern is concentration in real-system evidence: using the first listed D4a code as the primary substrate, 50 of 75 research entries are primarily evaluated on real clusters or hosts, compared with 15 simulation-first entries, 6 production-first entries, and 4 offline-first entries. At the same time, many papers triangulate evidence across modes: simulation appears somewhere in 21 of 75 entries, offline trace/model evidence appears somewhere in 8 of 75 entries, and emulation appears in only 2 of 75 entries. This supports a calibrated interpretation: real-system prototypes are common, but production deployment and explicit emulation remain comparatively sparse.

A second pattern is that workload/data realism is more common than explicit fidelity enhancement. Benchmark or trace-driven sources appear in 69 of 75 entries, while synthetic workload generation appears in 44 of 75 entries and often coexists with benchmark applications or trace replay. By contrast, explicit D4c fidelity enhancers are much rarer: fault injection, network shaping, or Kubernetes-in-the-loop evaluation appears in 11 of 75 entries. For dynamic breadth, the corpus shows that 51 of 75 entries model multiple D2 dynamic classes. The more persistent limitation is actuation and interaction breadth: 40 of 75 entries expose a single concrete D3b actuator, 26 expose multiple actuators, and 9 are predictive, diagnostic, or empirical studies without a concrete actuator in the coded row. This suggests that the central evaluation gap is not merely whether papers mention multiple dynamics, but whether they stress coupled dynamics, multiple actuators, and platform-mediated control-loop interactions in the same reproducible scenario.

{\tiny
\begin{table}[t]
\centering
\footnotesize
\setlength{\tabcolsep}{3.5pt}
\renewcommand{\arraystretch}{1.13}
\begin{tabular}{p{0.23\linewidth} p{0.28\linewidth} p{0.42\linewidth}}
\toprule
\textbf{Observed pattern} & \textbf{Count} & \textbf{Interpretation}\\
\midrule
Primary execution substrate (D4a; first listed code) &
\begin{tabular}[t]{@{}l@{}}
PD: 6/75 (8.0\%)\\
RC: 50/75 (66.7\%)\\
SM: 15/75 (20.0\%)\\
OF: 4/75 (5.3\%)
\end{tabular} &
Real-system testbeds remain the dominant primary evidence mode, while production-first evidence is sparse. Simulation-first and offline-first studies are important for design-space exploration, forecasting, and trace characterization, but should be interpreted differently from live controller validation.\\
\midrule
Auxiliary execution evidence in multi-label D4a cells &
\begin{tabular}[t]{@{}l@{}}
SM appears: 21/75 (28.0\%)\\
OF appears: 8/75 (10.7\%)\\
EM appears: 2/75 (2.7\%)
\end{tabular} &
Several papers combine real-system claims with simulation or offline analysis. These auxiliary codes should not be collapsed into primary-substrate counts, but they are useful for understanding how papers triangulate evidence.\\
\midrule
Workload or data source (D4b; multi-label) &
\begin{tabular}[t]{@{}l@{}}
SY: 44/75 (58.7\%)\\
BM: 43/75 (57.3\%)\\
TD: 37/75 (49.3\%)\\
BM or TD: 69/75 (92.0\%)
\end{tabular} &
Benchmark and trace-driven evidence is common, but synthetic workload generation is also widespread. Thus, workload realism should be assessed from the scenario matrix and replay/generation method, not from the benchmark or trace label alone.\\
\midrule
Explicit fidelity enhancer (D4c; FI/NE/KL) &
\begin{tabular}[t]{@{}l@{}}
Any FI/NE/KL: 11/75 (14.7\%)\\
FI: 9/75 (12.0\%)\\
NE: 4/75 (5.3\%)\\
KL: 1/75 (1.3\%)
\end{tabular} &
Fault injection is the most common explicit enhancer in the coded corpus, while explicit network shaping and Kubernetes-in-the-loop evaluation remain rare. This indicates that many real-cluster studies still leave orchestration and network dynamics only partially stressed.\\
\midrule
Modeled dynamic breadth (D2) &
\begin{tabular}[t]{@{}l@{}}
Single D2: 24/75 (32.0\%)\\
Multiple D2: 51/75 (68.0\%)
\end{tabular} &
The coding shows that many papers model more than one dynamic class. The remaining gap is therefore less about acknowledging multiple dynamics at all, and more about evaluating their coupled behavior under realistic control-loop interactions.\\
\midrule
Concrete actuation breadth (D3b) &
\begin{tabular}[t]{@{}l@{}}
No concrete actuator: 9/75 (12.0\%)\\
Single actuator: 40/75 (53.3\%)\\
Multiple actuators: 26/75 (34.7\%)
\end{tabular} &
More than half of the research entries evaluate one main actuation surface, and a non-trivial subset is predictive, diagnostic, or empirical rather than directly actuation-oriented. Multi-loop interaction remains less systematically evaluated than multi-dynamic modeling.\\
\bottomrule
\end{tabular}
\caption{Descriptive evaluation patterns observed in the coded research-paper subset of the representative system corpus (75 research-paper entries, excluding the 9 platform/documentation entries). Counts are descriptive statistics for the curated corpus rather than bibliometric prevalence estimates. For multi-label D4a cells, the first listed code is treated as the primary execution substrate; other multi-label rows report occurrence counts.}
\label{tab:eval-patterns}
\end{table}
}

These corpus-level observations motivate a more disciplined evaluation methodology. We therefore recommend the following principles for designing and reporting new systems.

First, comparison should be \emph{scenario-based}. Because cloud and microservice workloads are bursty, non-stationary, and highly sensitive to tail latency, steady-state averages alone are insufficient for judging adaptive controllers \cite{tail_at_scale2013,Bitbrains,Borg_trace}. A fair study should therefore define a scenario matrix that perturbs the dominant dynamics relevant to the proposal: workload bursts and diurnal shifts, request-mix or call-graph evolution, network-delay and bandwidth changes, co-location interference, and failure or degradation events when applicable. Each scenario should distinguish equilibrium behavior from transient behavior around change points, including how quickly the controller detects the shift, how far service quality degrades during adaptation, and whether recovery is monotonic or oscillatory.

Second, baseline selection should reflect both \emph{mechanism class} and \emph{operational realism}. If a paper proposes improved autoscaling, the comparison should include production-reactive baselines such as HPA/VPA and, where node elasticity matters, the Cluster Autoscaler \cite{K8s_HPA,k8s_vpa,k8s_cluster_autoscaler}. If the contribution lies in placement or traffic management, locality-unaware, network-aware, and graph-aware baselines should be included whenever those assumptions are relevant \cite{k8s_scheduling_CSUR2022,NetMARKS_InfoCOM21,deployment_routing_TPDS23,TraDE_TPDS2026}. When a system combines multiple actuators---for example, scaling plus routing or placement plus isolation---ablation studies are essential. Without ablation, the reader cannot determine whether a reported gain comes from better modeling, better decision logic, or simply from access to additional control knobs.

Third, reported metrics should capture \emph{control quality}, not just application-level efficiency. Percentile latency, throughput, error rate, utilization, cost, and energy remain necessary, but they are insufficient for dynamic environments \cite{deathStarBench_ASPLOS19,tail_at_scale2013,CarbonAware_2023}. Authors should also report the SLO-violation rate (or violation area), time to detection, time to recovery, action frequency, oscillation amplitude, service-placement churn (i.e., frequent re-placement decisions over time), and controller overhead, including decision latency, actuation latency, telemetry volume, and the controller's own resource consumption. These quantities are especially important when multiple control loops interact, because aggressive control can improve average performance while worsening stability or increasing operational overhead.

Fourth, strong claims should follow an \emph{evaluation ladder}. Simulation is useful for exploring large design spaces, emulation for reproducing network dynamics and controllable perturbations, and real clusters for validating end-to-end behavior under authentic control-plane timing \cite{CloudSim_SPE2011,Cloudnativesim_SPE25,Kollaps_EuroSys20,THUNDERSTORM_SRDS19}. Recent Kubernetes-in-the-loop systems and microservice emulation frameworks are especially valuable because they preserve realistic orchestration behavior without sacrificing experimental control \cite{k8sLoop_simu_PerformanceTools2024,chen2025idynamics}. However, when major claims rely on simulation or emulation, the modeled behavior should be calibrated against either a real deployment or published measurements; otherwise, improvements observed under idealized orchestration or static networks may not transfer to production settings.

Finally, reproducibility requires releasing the \emph{experimental context}, not just the controller source code. Public dataset surveys have shown that trace provenance, preprocessing decisions, and missing contextual information often determine whether a result can be replayed or compared fairly \cite{ACMDatasetsCloud2025}. At minimum, authors should document benchmark and trace versions, workload generators, Kubernetes and service-mesh versions, resource requests and limits, autoscaler settings, telemetry sampling policies, fault-injection procedures, random seeds, and deployment manifests or infrastructure-as-code whenever possible. Table~\ref{tab:repro-checklist} summarizes a practical checklist. In our view, two items deserve particular emphasis in future work: a clear \emph{dynamic scenario matrix} and an explicit \emph{baseline and ablation plan}. Together, they make it much easier to judge whether a reported improvement is genuinely robust to the kinds of dynamics that motivate the controller in the first place.

{
\scriptsize
\begin{table}[t]
\centering
\small
\renewcommand{\arraystretch}{1.12}
\begin{tabular}{p{0.25\linewidth} p{0.75\linewidth}}
\toprule
\textbf{Item to report} & \textbf{Rationale / examples}\\
\midrule
Dynamic scenario matrix & Workload bursts, request-mix or call-graph changes, network perturbations, interference, and failure injections; distinguish steady-state from transient phases.\\
Baselines and ablations & Reactive production baselines (e.g., HPA/VPA/Cluster Autoscaler), plus predictive or learning-based baselines when relevant; isolate the contribution of each actuator or model component.\\
Workload definition & Arrival process, request mix, burst patterns, replay rate, trace provenance, and any time compression or synthetic perturbation.\\
Microservice graph & Service count, fan-out, critical paths, and whether the dependency graph or communication pattern changes over time.\\
SLOs and metrics & p95/p99 latency, throughput, error rate, cost and energy, plus SLO-violation rate, recovery time, controller overhead, and oscillation or churn measures.\\
Control-loop details & Sampling period, decision latency, actuation latency, cooldown intervals, controller gains or hyperparameters, and stability safeguards.\\
System configuration & Kubernetes and service-mesh versions, node types, resource requests and limits, replica settings, and co-location assumptions.\\
Evaluation environment & Production vs.\ Real cluster vs.\ emulation/simulation, network shaping, interference injection, resource heterogeneity, and repeatability controls.\\
Artifacts & Code, configuration files, manifests, traces, containerized artifacts, and random seeds to enable exact replay of stochastic experiments.\\
\bottomrule
\end{tabular}
\caption{Reproducibility checklist for dynamics-aware microservice management studies.}
\label{tab:repro-checklist}
\end{table}

}

\section{Challenges and Future Directions}
\label{sec:challenges}

The taxonomy in Section~\ref{sec:taxonomy} reveals a literature rich in mechanisms yet fragmented in its treatment of coupled dynamics, uncertainty, and evaluation fidelity. In our view, the next stage of progress will depend less on introducing yet another isolated controller and more on building principled control stacks that connect telemetry, decision-making, and evaluation across the full D1--D4 design space. We therefore highlight six research directions that follow most directly from the gaps synthesized in this survey.

\subsection{Cross-layer Coordination, Timescale Separation, and Stability}
Dynamics-aware microservice management rarely operates through a single controller. Production stacks already expose a hierarchy of actuators---HPA/VPA, node autoscaling, service placement, service-mesh routing, admission control, and isolation---that operate at different scopes and timescales. The resulting interactions are not independent; rather, they form tightly coupled feedback loops. A routing change can shift hotspots across replicas and trigger autoscaling; autoscaling can alter request-queueing dynamics and service-placement pressure; node scaling and service re-placement can change locality, delay, and interference; and fault handling or rollback policies can abruptly invalidate controller assumptions \cite{k8s_cluster_autoscaler,k8s_vpa,k8s_interference_TCC,Dirigent_LC_QoS,istio,Envoy,collaborativeMesh_Infocom25}. As emphasized by the autonomic-computing paradigm and recent Kubernetes-oriented studies, these feedback loops arise naturally when multiple controllers observe overlapping symptoms but optimize different objectives or operate at different timescales \cite{autonomic_computing2003,k8s_scheduling_CSUR2022}.

A central research challenge is therefore to move from collections of locally reasonable controllers to \emph{composable control architectures}. Promising directions include explicit timescale separation, shared state estimation across layers (e.g., consistent views of demand, queueing state, and network latency), and coordination policies that reason about actuator precedence, cooldown windows, and rollback behavior. Control-theoretic analysis, queueing models, and system identification can help characterize stability margins, but they must be adapted to settings in which workload mix, service dependencies, and network state evolve online. Recent verification work such as Kivi makes this agenda more concrete by checking whether interacting Kubernetes controllers and event interleavings can violate user intent, showing that controller coupling is a verification problem as well as a tuning problem \cite{Kivi_arxiv23}. Equally important, future work should report controller-interaction effects directly---for example, whether gains from routing persist once autoscaling or node provisioning is enabled. The long-term goal is not a single monolithic controller, but principled multi-loop coordination that preserves stability, avoids oscillations, and remains interpretable to operators.

\subsection{Telemetry-to-Control under Partial Observability, Causality, and Uncertainty}
Observability stacks have made metrics, logs, traces, and service-mesh signals widely available, but availability does not automatically imply control utility. Metrics aggregate multiple causes into coarse-grained time series; traces are sampled and may miss critical spans; topology changes alter the meaning of historical baselines; and queueing, contention, or partial failures often remain only indirectly observable \cite{Prometheus,opentelemetry,dapper_tracing2010,jaeger,zipkin,XTrace_NSDI07,PivotTracing_SOSP15,TProf_SoCC21,STEAM_FSE23,DeepFlow_SIGCOMM23,Sage_ASPLOS21,Seer,PerfScope_bug_inference,ART_ASE24,RootCauseAnalysis_CSUR_2025}. In other words, microservice control operates under a persistent observability gap: the controller must act on a delayed and incomplete projection of the true system state. This challenge is particularly acute because the same symptom---for example, an increase in p99 latency---may be caused by workload bursts, downstream congestion, call-graph changes, noisy-neighbor interference, or a partially degraded dependency.

The research problem is therefore not simply better monitoring, but \emph{control-oriented inference}. Future systems require telemetry representations that are compact enough for online control yet expressive enough to preserve causal structure across services, resources, and network paths. Promising directions include uncertainty-aware state estimation, counterfactual and causal models for dependency reasoning, adaptive tracing and active probing, and controllers that optimize under uncertainty (e.g., using confidence intervals) rather than relying on point estimates alone. Another open question concerns how uncertainty should be surfaced to human operators: if a controller recommends scaling, rerouting, or remediation, it should also communicate whether that action is supported by strong evidence or by ambiguous correlations. Bridging observability and control in this way is essential for enabling dynamics-aware management to remain both robust and trustworthy under partial observability.

\subsection{Coupled Multi-Dynamics and Multi-Objective Control}
Many existing approaches now acknowledge more than one source of variation, but they still tend to cover only a limited slice of the coupled dynamics that production microservices exhibit. As Table~\ref{tab:eval-patterns} shows, multiple D2 dynamic classes appear in 51 of 75 coded research entries; however, many of these combinations remain local, such as workload plus dependency, network plus placement, or contention plus resource shaping. Production systems couple these dynamics more broadly: request-mix shifts alter critical paths, deployment changes reshape traffic matrices, network delays shift effective bottlenecks, and co-location interference propagates into tail latency and cost \cite{SoCC2021_MS_luo,metaTraces_ATC23,NetMARKS_InfoCOM21,deployment_routing_TPDS23,TraDE_TPDS2026,Cloud_White}. At the objective layer, a single actuation decision can simultaneously affect latency, throughput, resource cost, energy consumption, and, in some cases, carbon exposure \cite{cost_efficient_MS,sharedMS_AcmTranCS,CarbonAware_2023,saxena2022proactive,Hipster_LC_task_manager,Rubik_LC_powerManagement}. The difficulty lies not only in the multiplicity of objectives, but also in the fact that both system dynamics and the trade-offs among objectives evolve over time.

This motivates a stronger agenda centered on \emph{joint models and joint decision-making}. Future controllers should reason jointly about workload, call-graph, network, and interference signals, rather than treating them as separate optimization layers connected only through heuristics. They also require objective formulations that make trade-offs explicit: for example, when temporary relaxation of SLO targets is acceptable to reduce carbon footprint, or when cost constraints should yield to resilience during failure episodes. Recent carbon-aware systems for web services, such as CASPER, demonstrate that SLO-aware carbon optimization is feasible, while contemporaneous work on spatiotemporal workload shifting underscores that the attainable gains depend on workload characteristics and latency tolerance \cite{CASPER_IGSC23,CarbonShiftLimit_EuroSys24}. Promising directions include hierarchical multi-objective optimization, constrained control with dynamic budgets, and online decomposition that separates fast operational objectives from slower economic or sustainability objectives. A key challenge is determining which couplings must be modeled explicitly and which can be abstracted without degrading control quality. The field, therefore, needs methods that are expressive enough to capture cross-dynamics interactions while remaining identifiable, stable, and interpretable in deployment.

\subsection{Safe, Sample-Efficient, and Trustworthy Learning-Based Controllers}
Learning-based controllers are attractive because microservice management involves large action spaces, non-linear response surfaces, and non-stationary workloads. RL-based systems such as ConAdapter, CoScal, and SafeDRL demonstrate that adaptive policies can coordinate multiple resource and scaling knobs in response to changing demand \cite{ms_rl_SOCC23,ms_rl_TNSM,SafeDRL_TC24}. Yet these same strengths also make online deployment risky: continuous adaptation and automation can amplify errors in real time. Exploration incurs real performance costs; rare failures are difficult to learn from safely; policies trained under one workload mix or dependency graph may generalize poorly when service topology, co-location patterns, or network conditions change; and reward design can obscure undesirable behaviors such as oscillations, excessive actuation, or unfairness across services. Related evidence from cluster scheduling for ML workloads reinforces these concerns, particularly with respect to fairness, robustness, and distribution shift \cite{Pollux_ClusterScheduling_goodput,INFOCOM_online_job_scheduling,ONES_GPUscheduling}.

Accordingly, the next step is not simply ``more RL,'' but \emph{trustworthy learning for cloud control}. Promising directions include constrained policies and safety-shielded RL, hybrid designs that warm-start from analytical models or expert policies, offline or trace-driven pretraining followed by limited online adaptation, and explicit uncertainty estimation for detecting out-of-distribution states. Another critical direction is evaluation: learning-based controllers should be stress-tested under diverse dynamic scenario matrices, with transparent reporting of instability episodes, rollback frequency, and worst-case SLO violations, rather than only average reward \cite{k8sLoop_simu_PerformanceTools2024,chen2025idynamics}. Human override and auditability of policy decisions are also essential for production deployment. In practice, a promising design pattern is one that operates within clearly specified safety envelopes and gracefully degrades to conservative fallback policies when confidence is low.

\subsection{Standardized Dynamic Scenarios, Benchmarks, and Reproducible Evaluation Pipelines}
Our coded corpus suggests that evaluation remains uneven, especially for coupled dynamics and multi-actuator settings. Benchmark suites such as DeathStarBench, $\mu$Bench, and TrainTicket improve comparability, while production traces, graph-generation methods, network emulators, and Kubernetes-in-the-loop frameworks capture complementary aspects of realism \cite{deathStarBench_ASPLOS19,muBench_TPDS23,trainTicket,SoCC2021_MS_luo,metaTraces_ATC23,GrahGeneration_ICS25,Kollaps_EuroSys20,THUNDERSTORM_SRDS19,k8sLoop_simu_PerformanceTools2024,chen2025idynamics}. Yet Table~\ref{tab:eval-patterns} shows a gap between workload/data realism and explicit fidelity enhancement: benchmark or trace-driven sources appear in 69 of 75 research entries, whereas explicit D4c fidelity enhancers appear in only 11 of 75. To our knowledge, the field still lacks widely accepted, standardized \emph{dynamic scenario} suites that combine these components reproducibly. Many published studies use benchmarks or traces without a full scenario matrix, leaving it unclear whether reported gains persist under call-graph evolution, transient failures, network perturbations, multi-tenant interference, or interacting actuators. As discussed in Section~\ref{subsec:eval-repro}, what matters is not merely the benchmark name, but the scenario matrix, the actuation context, and the fidelity of the control plane.

Future progress depends on treating evaluation as shared infrastructure rather than as paper-specific glue code. The field would benefit from curated scenario suites that parameterize bursts, request-mix shifts, routing changes, placement churn, failure injection, and interference episodes; standardized packaging formats for traces, manifests, workload drivers, and telemetry schemas; and stronger norms for releasing controller configurations, random seeds, and replay scripts \cite{ACMDatasetsCloud2025,CloudSim_SPE2011,Cloudnativesim_SPE25}. Another promising direction is to couple trace-driven workloads with Kubernetes-in-the-loop or emulation-based evaluation so that controller decisions experience realistic orchestration delays and network behavior without sacrificing repeatability. In short, the goal is not merely more benchmarks, but a reproducible evaluation pipeline that enables systematic comparison across laboratories, control mechanisms, and the diverse dynamics that motivate this field.

\subsection{Agentic and Multi-Agent LLMs for Dynamics-Aware Management}
Recent LLM-era AIOps work positions incident understanding, runbook execution, and operational decision support as a frontier for autonomous cloud operations \cite{Zhang2025AIOpsSurvey}. AIOpsLab frames this direction as \emph{AgentOps}, with microservice deployment, fault injection, workload generation, telemetry export, and agent--cloud interfaces for evaluation \cite{AiAgentCloud_AIOps_SoCC24,AIOpsLab_MLSys25}. However, dependable autonomy remains immature: ITBench reports that strong contemporary agents solve only 11.4\% of its SRE (Site Reliability Engineering) scenarios \cite{ITBench_ICML25}. Representative systems illustrate the emerging design pattern: multi-agent diagnosis and remediation for Kubernetes configuration errors, modular supervisor-coordinated Kubernetes management, rollback-aware autonomous SRE, and memory-augmented failure diagnosis \cite{KubeLLM_CloudSummit25,KubeIntellect_arxiv25,STRATUS_NIPS25,MetaKube_arxiv26}.

For dynamics-aware microservice management, agentic LLMs are better viewed as a slower supervisory layer than as replacements for fast autoscaling, scheduling, or service-mesh control loops. Their plausible role is to integrate telemetry, topology, configuration, and policy context into an interpretable \emph{world model}; identify dominant D2 dynamics; coordinate D3 actuators across D1 scopes; and produce operator-facing explanations or mitigation plans. Credible deployment requires grounding decisions in runtime state, enforcing safety envelopes for high-impact actions (policy checks, budgets, rollouts, rollbacks, and human approval), curating incident memory to avoid stale remediation, and evaluating agents under replayable microservice dynamics with safety monitoring. Thus, the value of agentic and multi-agent LLMs will depend less on language-model capability alone than on integration with existing control primitives, observability stacks, safety mechanisms, and D4 evaluation fidelity \cite{AIOpsLab_MLSys25,ITBench_ICML25,STRATUS_NIPS25}.

\section{Summary and Conclusions}
\label{sec:conclusion}

Microservice management is often poorly captured by purely steady-state formulations.
In production environments, demand varies, request paths evolve, resources are shared, and network and infrastructure conditions fluctuate. These dynamics interact with cloud-native control and data planes, making autoscaling, placement, routing, isolation, and remediation frequently coupled in production settings rather than independent decisions. The central premise of this survey is therefore that adaptive microservice management is better designed as a coordinated control stack under dynamics, rather than as a collection of isolated mechanisms.

To organize this space, we proposed a taxonomy along four primary dimensions: where control is applied (D1), which dynamics are modeled (D2), how adaptation is performed (D3), and how systems are evaluated (D4), together with two orthogonal dimensions---objectives/SLOs and telemetry. This taxonomy helps separate concerns that are often conflated in prior literature. In particular, it clarifies that the same actuator can address different dynamics, that the same dynamic can be addressed by multiple mechanisms, and that evaluation methodology is not an afterthought but a major contributor to the credibility and interpretability of reported results. Using this structure, we mapped representative systems, benchmark suites, traces, emulators, and simulation tools, and showed how the literature spans predictive control, graph-aware management, network-aware scheduling, interference mitigation, diagnosis and remediation, sustainability-aware designs, and emerging agentic AI-assisted operations.

Several broader lessons emerge from this synthesis. First, while many papers now encode more than one dynamics class, the field still lacks systematic evaluation of coupled dynamics under multiple interacting actuators and realistic control-plane behavior. Second, moving upward in abstraction---from hosts and containers to services and application graphs---improves semantic alignment with end-to-end SLOs, but also increases telemetry requirements and reliance on accurate dependency models. Third, no adaptation mechanism is universally superior: prediction, scaling, routing, isolation, learning, and remediation each exhibit distinct strengths and failure modes, and their interactions often matter more than individual algorithms considered in isolation. Fourth, many reported gains are sensitive to the evaluation environment. Results obtained under simplified workloads or static networks can change substantially once realistic orchestration delays, call-graph variation, and multi-tenant interference are introduced.

For researchers, this survey suggests that the next phase of progress will come from \emph{coordination across axes}: cross-layer controller design, stronger telemetry-to-control abstractions under partial observability, safe and interpretable learning-based control, emerging agentic LLM supervision, and reproducible evaluation that better reflects real operational dynamics. For practitioners, the main implication is equally direct: production-ready microservice management should be designed as a coordinated control stack, in which observability, objectives, actuators, and safety mechanisms are specified jointly rather than assembled independently. We hope this survey provides a useful foundation for this agenda and helps move the field toward cloud-native systems that are not only more efficient but also more robust, interpretable, and reproducible.

\bibliographystyle{ACM-Reference-Format}
\bibliography{references}

\end{document}